%% file: neurips_2025.tex
\documentclass{article}

\usepackage[preprint]{neurips_2025}

\usepackage[utf8]{inputenc} 
\usepackage[T1]{fontenc}    
\usepackage{hyperref}       
\usepackage{graphicx}       
\usepackage{url}            
\usepackage{booktabs}       
\usepackage{amsfonts}       
\usepackage{nicefrac}       
\usepackage{microtype}      
\usepackage{xcolor}         
\usepackage{multirow}
\usepackage{tikz}
\usepackage{amsmath}
\usepackage{array}
\usepackage{wrapfig}
\usepackage{subcaption}
\usepackage{xspace}
\usepackage{placeins}
\usepackage{comment}  
\usepackage{cleveref}
\usepackage{float}

\newcommand{\algo}{\texttt{PnP}\xspace}
\newcommand{\algofull}{Plug-and-Plant (\algo)\xspace}

\newcommand{\ct}[1]{\raisebox{.5pt}{\textcircled{\raisebox{-.9pt} {#1}}}}

\newif\ifcomments
\commentstrue

\ifcomments
\newcommand{\authorcomment}[2]{\tikz[baseline=(X.base)]\node [draw=#1,fill=#1!40,semithick,rectangle,inner sep=2pt, rounded corners=3pt] (X) {#2};}

\newcommand{\cz}[1]{\authorcomment{blue}{Chaoyi:} \textcolor{blue}{\textit{#1}}}
\newcommand{\lc}[1]{\authorcomment{red}{Lydia:} \textcolor{red}{\textit{#1}}}
\newcommand{\py}[1]{\authorcomment{orange}{Pinyu:} \textcolor{orange}{\textit{#1}}}

\newcommand{\ry}[1]{\authorcomment{green}{Renyi:} \textcolor{green}{\textit{#1}}}
\newcommand{\rb}[1]{\authorcomment{cyan}{Robert:} \textcolor{cyan}{\textit{#1}}}
\else
\newcommand{\authorcomment}[2]{}

\newcommand{\cz}[1]{}
\newcommand{\lc}[1]{}
\newcommand{\py}[1]{}
\newcommand{\ry}[1]{}
\newcommand{\rb}[1]{}
\fi

\title{Optimization-Free Universal Watermark Forgery with Regenerative Diffusion Models}

\author{Chaoyi Zhu\thanks{Equal Contribution} $^{1}$, ZAITANG LI$^{*2}$, Renyi Yang$^1$, \\ \bf Robert Birke$^4$, Pin-Yu Chen$^5$, Tsung-Yi Ho$^2$, Lydia Y. Chen\thanks{Corresponding Authors} $^{1,3}$\\
$^1$Delft University of Technology, $^2$The Chinese University of Hong Kong\\
$^3$University of Neuchâtel, $^4$University of Turin, $^5$IBM Research\\
\texttt{\{c.zhu-2, y.chen-10\}@tudelft.nl}, \texttt{r.yang-7@student.tudelft.nl}\\
\texttt{\{ztli, tyho\}@cse.cuhk.edu.hk}, \texttt{robert.birke@unito.it}, \texttt{pin-yu.chen@ibm.com}\\
\And
}

\begin{document}

\maketitle

\begin{abstract}
  Watermarking becomes one of the pivotal solutions to trace and verify the origin of synthetic images generated by artificial intelligence models, but it is not free of risks. Recent studies demonstrate the capability to forge watermarks from a target image onto cover images via adversarial optimization without knowledge of the target generative model and watermark schemes. In this paper, we uncover a greater risk of an optimization-free and universal watermark forgery that harnesses existing regenerative diffusion models. Our proposed forgery attack, \algo (Plug-and-Plant), seamlessly extracts and integrates the target watermark via regenerating the image, without needing any additional optimization routine. It allows for universal watermark forgery that works independently of the target image’s origin or the watermarking model used. We explore the watermarked latent extracted from the target image and visual-textual context of cover images as prior to guide sampling of the regenerative process. Extensive evaluation on 24 scenarios of model-data-watermark combinations demonstrates that \algo can successfully forge the watermark (up to 100\% detectability and user attribution), and maintain the best visual perception.  By bypassing model retraining and enabling adaptability to any image, our approach significantly broadens the scope of forgery attacks, presenting a greater challenge to the security of current watermarking techniques for diffusion models and the authority of watermarking schemes in synthetic data generation and governance. Our code is available at the repository: \url{https://github.com/chaoyitud/PnP-Watermark-Forgery}.

\end{abstract}

\input{Sections/Introduction}

\input{Sections/RelatedWorks_v2}

\input{Sections/Method}
\input{Sections/Experiments}
\input{Sections/Conclusion}

\bibliography{main.bib}
\newpage
\appendix

\newpage
\input{Sections/Impl_Hardware_Details}

\input{Sections/Quality_Table}

\input{Sections/Appendix-addtional-examples}

\end{document}

%% file: Sections/Introduction.tex
\section{Introduction}
The rapid advancement of generative models~\cite{10521640} has led to a significant increase in the creation of high-quality images.  As synthetic images proliferate, tracking their provenance~\cite{bing,synthid} is becoming ever crucial. 
Tracing their usages and verifying their origins are key steps, especially in case of misuse or misconduct. As a result, the demand for effective watermarking techniques to protect intellectual property and ensure the authenticity of generated content has become increasingly important.

Watermarks are a promising solution for embedding information into images, enabling the identification of the creator or source of the content. Post-processing watermarks~\cite{PereiraP00,BiSHYH07} studied for decades add a covert marker into images as noise on the pixels, but impact visual quality and have low resilience to post-editing attacks.
Semantic watermarking~\cite{jiang2024watermarkbasedattributionaigeneratedcontent} overcomes such limits by adding the watermark at the latent space of synthetic data, often during the sampling/generation process. For instance, TreeRing~\cite{treerings2023} and Gaussian Shading~\cite{yang2024gaussian} are watermarks for images which are embedded at the initial noise of the diffusion process. This allows the watermark to maintain minimum impact to the image quality, while remaining detectable under various post-editing attacks, e.g., paraphrasing and translation attacks. Furthermore, semantic watermarks can be straightforwardly integrated into existing diffusion models.

\begin{figure}
    \centering
    \includegraphics[width=\linewidth]{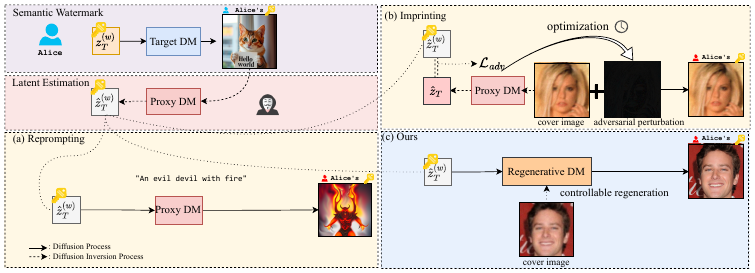}
    \caption{Comparison of different Watermark forgery methods for diffusion models (DM): (a) Reprompt forges the watermarks on text-to-image diffusion models; (b) Imprint forges watermarks on cover images using adversarial optimization on the proxy DM; and (c) \algo, our proposed optimization-free forgery uses controllable regenerative DM. }
    
    \label{fig:intro_example}
\end{figure}

Forgery is one of the long standing challenges for data governance, where adversaries can falsify the data without legit authorization. Recent studies~\cite{mueller2024forgery} demonstrate that the aforementioned semantic watermark for synthetic images can be extracted and forged onto different images by using proxy diffusion models.  
Reprompt~\cite{mueller2024forgery}, see Fig.~\ref{fig:intro_example}(a), leverages a proxy diffusion model, specifically text-to-image, and uses different prompts and the extracted ``implicit'' watermark to generate new images. Imprint~\cite{mueller2024forgery}, see Fig.~\ref{fig:intro_example}(b),  demonstrates a stronger forgery attack on general diffusion models by explicitly embedding the watermark through adversarial optimization. However, reprompting fails to forge watermarks across any images, while imprinting produces watermarked outputs with notable quality degradation and incurs long optimization time, rendering it impractical for large-scale use.

In this work, we unveil  
\algofull, a watermark forgery attack
for images that are generated by diffusion models with either semantic watermarks. 
Our key proposition is to leverage existing regenerative diffusion models, such as super-resolution~\cite{chen2310pixart,wang2024exploiting}, face restoration~\cite{lin2024diffbir}, or watermark removal~\cite{liu2024image}, as a backbone for the forgery, without needing any additional finetuning, see Fig.~\ref{fig:intro_example}(c).
\algo allows for universal watermark forgery that works independently of the source image’s origin or the watermarking model used. \algo is a light-weight forgery method that uses a controllable regenerative diffusion model 
to forge the watermark onto either synthetic or real-world images. 
Specifically, we first extract the watermarked latent from the target image and then explore such latent as prior to regenerate the cover image under its visual-textual guidance as constraints. The forged watermark and key characteristics of cover images are preserved in the generated images. 
Thus, 
\algo has 
two-fold advantages. First, as we are an optimization-free solution, the forged images can be generated up to 50 times faster than existing methods. Secondly, the regenerative diffusion model allows for improved image resolution and quality, overcoming the quality degradation typically associated with adversarial optimization-based methods.
    
Our findings underline the ease in forging watermarks with off-the-shelf generative models. This allows even users with no forgery expertise to launch such attacks. Hence, model owners cannot reliably depend on current watermarking approaches for data governance, suggesting demands on more advanced and forgery-proof watermarking methods. Our specific technical contributions are:
\begin{itemize}
    \item We propose, \algo, a plug-and-play universal watermark forgery attack that forges the watermark of a target image onto any arbitrary cover image, by harnessing regenerative diffusion models.
    \item \algo extracts the forged watermark latents and exploits both the forged latents and visual-textual contexts to condition the regeneration of cover images with the forged watermarks.
    \item Extensive experimental validation,  across multiple state-of-the-art watermarking techniques and regenerative diffusion models, demonstrates that, \algo not only forges watermarks successfully but also improves the overall image perceptuality. 

\end{itemize}

%% file: Sections/RelatedWorks_v2.tex
\section{Background and Related Studies}

\subsection{Image Watermarking Methods}

Image watermarking methods can be broadly classified into two categories: \textit{post-processing} and \textit{in-processing} methods. {Post-processing watermarks} modify the final output, embedding the watermark in the image content~\cite{PereiraP00,BiSHYH07}. {In-processing watermarks}, on the contrary, 
are applied during image generation and embed the watermark in the model's internal representation.
A unique subcategory of in-processing methods is \textit{semantic watermarking}. In this approach, the watermark is embedded directly into the initial latent space of the image. This latent representation carries the watermark through the image generation process, ensuring its persistence. Moreover, the watermark can be recovered from the generated image using a diffusion inversion process.
We focus on two popular semantic watermarks: Tree-Ring~\cite{treerings2023} and Gaussian Shading~\cite{yang2024gaussian}.

\textbf{Tree-Ring} watermarking is a pioneering semantic watermarking method that embeds a unique pattern into the Fourier domain of the initial latent space. This pattern consists of concentric rings arranged in a tree-like structure. 
Watermark detection involves the diffusion inversion process, followed by evaluating the distance between the recovered initial latent space and the predefined tree-ring structure in the Fourier domain. An image is watermarked if the distance falls below a model-specific threshold
\textbf{Gaussian Shading} watermarking embeds a secret code by manipulating the quadrants of the initial latents within the Gaussian distribution. During detection, the watermark is recovered by inverting the diffusion process and checking whether the given value lies within the designated quadrant. Since the watermarked latent follows a Gaussian distribution, the watermarking method achieves near-lossless image quality for a single image. Our forgery attack is effective against both methods.



\subsection{Watermark Forgery Attacks}
While several related works address watermark forgery attacks, most focus on Tree Ring and do not extend to Gaussian Shading. Currently, only \cite{mueller2024forgery} can successfully forge Gaussian Shading watermarks via two novel attacks: \textit{Reprompt} and \textit{Imprint}.
Both attacks use DDIM inversion via a proxy diffusion model to obtain the initial latent representation (see Fig.~\ref{fig:intro_example}). In \textit{Reprompt}, a new image is generated by the diffusion model using the recovered initial latent and a different prompt. This attack is limited to forging watermarks on synthetic images generated by the diffusion model. In \textit{Imprint}, the adversary uses the target image as cover and introduces adversarial noise, which is then optimized through the diffusion process to recover a latent representation similar to that of the watermarked image. Due to the multiple diffusion steps, this optimization process is both computationally intensive and time-consuming. Additionally, the adversarial noise can degrade the quality of the cover image. 

Different from previous attacks, we leverage a regenerative diffusion model to embed the watermark into the target cover at low computational cost and superior perceived quality. Moreover, as a side benefit, our method can remove  post-processing watermarks from original images before forging the target semantic watermark.

\subsection{Regenerative Diffusion Models}

Regenerative diffusion models have emerged as a transformative approach for enhancing visual content, with significant advancements in image super resolution and restoration. \cite{wang2024exploiting} presents a framework to exploit the diffusion prior for real world image super resolution. It employs a time-aware encoder and feature warping to balance fidelity and perceptual quality 
by aligning the restoration process with the generative distribution of pre-trained models. \cite{lin2024diffbir} introduces DiffBIR, a general restoration pipeline for blind image restoration. By decoupling the process into degradation removal and information regeneration, 
it enables robust recovery of fine-grained details in degraded images.

Semantic-aware super resolution refers to the task of enhancing low-resolution images while integrating high-level semantic information  to guide the reconstruction process, ensuring that the upscaled outputs maintain semantic consistency and structural plausibility.
For this task, \cite{wu2024seesr} proposes SEESR, which trains a degradation-aware prompt extractor to generate soft and hard semantic prompts; by integrating low-resolution (LR) images into the initial sampling noise, it mitigates the diffusion model’s tendency to generate excessive random details, preserving semantic fidelity. \cite{sun2401improving} focuses on improving the stability of diffusion models for content-consistent super resolution, achieving consistent content across scales through refined latent space manipulation. HoliSDiP~\cite{tsao2024holisdip} advances this field by combining holistic semantics and diffusion prior, using semantic labels as text prompts and segmentation masks for dense guidance to enhance image quality in real world scenarios.
These works collectively highlight the versatility of regenerative diffusion models, while also paving the way for addressing remaining challenges in balancing controllability and generative fidelity for more complex real-world applications. 

While existing approaches predominantly concentrate on restoration and semantic guidance, SuperMark~\cite{hu2024supermarkrobusttrainingfreeimage} demonstrates the application of regenerative diffusion models to watermarking tasks. We extend the utility of regenerative diffusion models to watermark forgery, exploiting their inherent latent space manipulation capabilities to achieve seamless embedding of target watermarks into arbitrary images through optimization-free methods.

%% file: Sections/Method.tex
\section{Method - \algo Forgery Attack}
\label{sec:method}
\begin{figure}[t]
    \centering
    \includegraphics[width=0.99\linewidth]{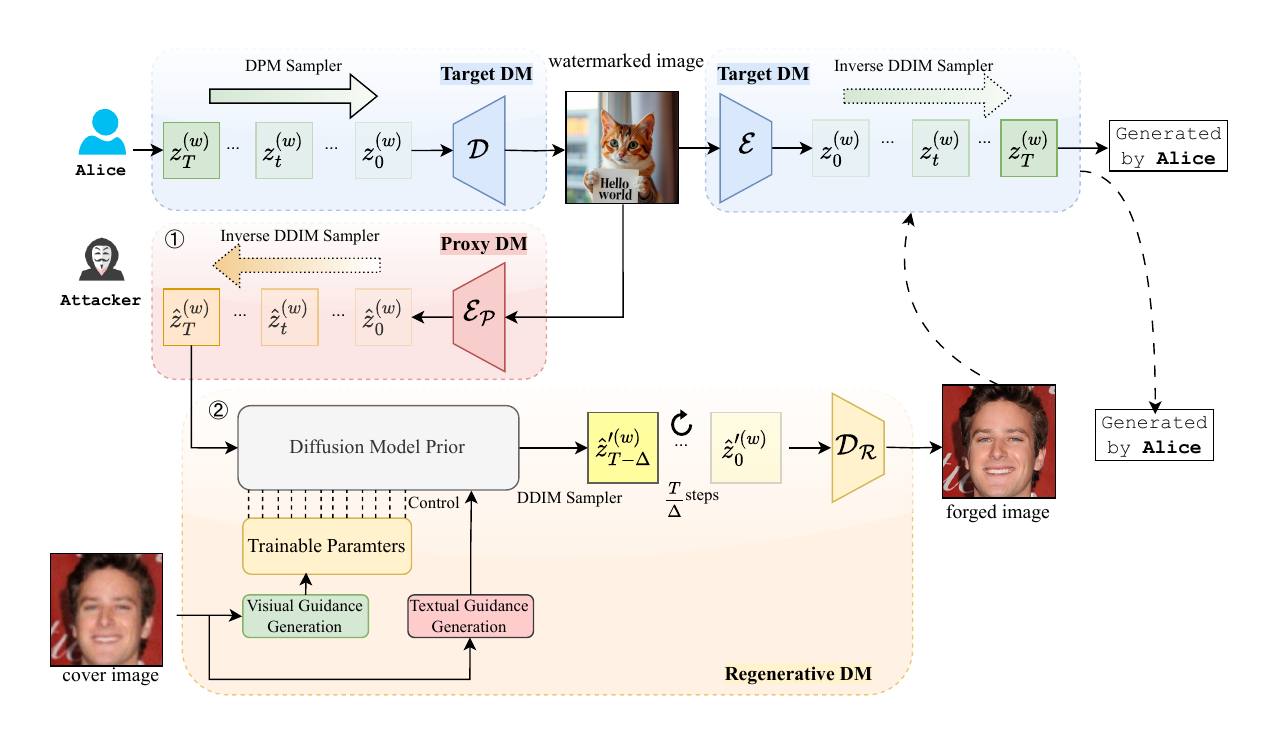}
    \caption{Overview of our watermark forgery approach: \algo. The process consists of two main stages: \ct{1} Watermark Latent Estimation and \ct{2} Regeneration with Regenerative diffusion models. In the first stage, we estimate the latent representation of the target watermark using a black-box approach based on a publicly available proxy diffusion model. In the second stage, we leverage regenerative diffusion models to manipulate the watermark in a target image without access to the original model parameters.}
    \label{fig:enter-label}
\end{figure}

In this section, we present \algo, our novel approach for watermark forgery, which leverages regenerative diffusion models to manipulate semantic watermarks without access to the original diffusion model parameters. Our approach, illustrated in Fig.~\ref{fig:enter-label}, consists of two key stages: \ct{1} watermark latent estimation highlighted via the red box; and \ct{2} regeneration with regenerative diffusion models enclosed in the orange box. Notice that Stage \ct{1} extracts the key characteristics of the watermark ensuring that it is embedded in any image outputted by Stage \ct{2}. This stage is only needed once for each watermark to be forged.
Once extracted, different cover images provide different new semantic features for the images to be generated by the regenerative diffusion model in Stage \ct{2}. In our example, the cat image contains the watermark of Alice, which we want to embed into the cover profile picture. After regeneration, watermark detection attributes both images to Alice.

\textbf{Adversarial assumptions}.
Here, we assume adversaries are curious and have no knowledge about the target model, except being a diffusion model, nor the watermarking method. Adversaries have access to proxy diffusion models and computation capacity to 
perform inference on the proxy and regeneration diffusion models. No adversarial-optimization or finetuning is required.

\subsection{Watermark Latent Estimation}
\label{sec:estimation}
Our approach begins in Stage \ct{1} by estimating the latent representation of the target watermark using a black-box approach based on the idea from~\cite{mueller2024forgery}. Semantic watermarking systems assume security through the secrecy of the diffusion models. However, existing work has shown that watermarking models can be attacked in a black-box setting, where the attacker does not have access to the model parameters or training data.

Denote the target diffusion model as $\Theta = (\mathcal{E}, \mathcal{U}, \mathcal{D})$, where $\mathcal{E}$ is the encoder, $\mathcal{U}$ is the noise predictor, and $\mathcal{D}$ is the decoder. The semantic watermark is typically embedded in the latent noise $z_T^{(w)}$ at the initial time step $T$, which is then denoised and decoded to produce the final watermarked image $x^{(w)}$.

Take a publicly available diffusion model as the proxy model $\Theta_P = (\mathcal{E}_P, \mathcal{U}_P, \mathcal{D}_P)$. It can be used to forge watermarks without access to $\Theta$. Our approach exploits the shared latent space structure between diffusion models to estimate the watermark's latent representation. Given a watermarked target image $x^{(w)}$ generated by the target model $\Theta$, our goal is to estimate the latent noise $\hat{z}_T^{(w)}$ embedded with the watermark. The procedure is as follows:

\begin{enumerate}
\item \textbf{Latent Encoding:} The attacker uses the proxy encoder $\mathcal{E}_P$ to map the watermarked image $x^{(w)}$ to the proxy model's latent space:
\begin{equation}
\hat{z}_0^{(w)} = \mathcal{E}_P(x^{(w)}).
\end{equation}
Given a watermarked target image $x^{(w)}$ generated by the target model $\Theta$, our goal is to estimate the latent noise $\hat{z}_T^{(w)}$ that encodes the watermark. 
\item \textbf{Noise Estimation:} The inverse DDIM sampler $\mathcal{I}_{0 \to T}$ of the proxy model is applied to estimate the latent noise $\hat{z}_T^{(w)}$ from $\hat{z}_0^{(w)}$:

\begin{equation}
    \hat{z}_T^{(w)} = \mathcal{I}_{0 \to T}(\hat{z}_0^{(w)}; \mathcal{U}_P).
\end{equation}

\end{enumerate}

This estimated noise $\hat{z}_T^{(w)}$ captures the semantic information embedded in the watermark, which we leverage in the subsequent stage to forge watermarks.

\subsection{Regeneration with Regenerative Diffusion Models}  
In Stage \ct{2} of our watermark forgery approach, as illustrated in Fig.~\ref{fig:enter-label}, we leverage regenerative diffusion models to manipulate the watermark in a target image without access to the original model parameters. This stage is a crucial part of our framework, consisting of several key components and processes that work together to achieve watermark forgery.  

\subsubsection{Watermark Latent Estimation Utilization}  
In the first stage of watermark latent estimation, we utilize this latent representation in the regenerative process. As shown in the figure, the attacker first uses the inverse DDIM sampler of the proxy diffusion model (\(\Theta_P\)) to estimate the latent noise. This \(\hat{z}_T^{(w)}\) captures the semantic information embedded in the watermark, serving as input latents for the regenerative diffusion model.  

\subsubsection{Regenerative Diffusion Model Architecture and Process}  

The regenerative diffusion model at this stage integrates a pretrained, publicly available text-to-image (T2I) diffusion model prior \(\mathcal{U}_{R}\), with its parameters kept fixed. To enable watermark regeneration, a set of lightweight, trainable components \(\mathcal{F}_{R}\) is introduced. The cover image \(x_{c}\) is first processed by an image encoder \(\mathcal{E}_{R}\) to extract visual embeddings \(\mathcal{E}_{R}(x_{c})\), which are then passed to the trainable modules. In some variants, large language models (LLMs) or captioning models \(\mathcal{T}_{R}\) are employed to extract textual information \(\mathcal{T}_{R}(x_{c})\), providing prompt-based textual guidance to condition the diffusion prior.

During training, the regenerative model aims to reconstruct the cover image at the same or higher resolution by conditioning the generative process on both the visual and textual signals extracted from \(x_c\). The learned modules bridge these signals and guide the generation process, ensuring the preservation of the original image content in a visually consistent manner.

Specifically, the DDIM sampler \(\mathcal{G}\) operates over \(\frac{T}{\Delta}\) steps, where \(T\) denotes the number of sampling steps of the target diffusion model, and \(\Delta\) is the step length used in the regenerative model. It transforms the estimated watermark latent \(\hat{z}_{T}^{(w)}\) into a regenerated latent representation \(\hat{z}_0^{\prime(w)}\), conditioned on both visual and textual features:

\begin{equation}
    \hat{z}_0^{\prime(w)} = \mathcal{G}_{T \to 0} \left(\hat{z}_{T}^{(w)} \middle| \mathcal{E}_{R}(x_{c}), \mathcal{T}_{R}(x_{c})\\;\, \mathcal{U}_R, \mathcal{F}_{R} \right),
\end{equation}

where \(\mathcal{E}(x_c)\) and \(\mathcal{L}(x_c)\) provide the visual and textual conditions, respectively. The decoder \(\mathcal{D}_R\) then maps the refined latent back to the image space to generate the forged image:

\begin{equation}
    x_c^{(w)} = \mathcal{D}_R\left( \hat{z}_0^{\prime(w)} \right).
\end{equation}

\subsubsection{Transferability and Forgery Realization}  
The design of the regenerative diffusion model ensures transferability to the target model. By exploiting the shared latent-space structure and diffusion priors across different diffusion models, the regenerative process can effectively manipulate the watermark. 

During inference, the combination of visual and textual guidance, along with trainable parameters, steers the denoising trajectory. This adaptation allows the model to regenerate the cover image in a consistent manner, while the preserved diffusion prior and similar denoising process help retain the semantic information of the target watermark.

Specifically, by injecting the estimated watermark latent $\hat{z}_T^{(w)}$ into the regenerative diffusion model and using the associated control mechanisms, such as visual and textual guidance, we can synthesize the appearance of a watermark from the original target model $\Theta$ into an arbitrary cover image. This approach operates in a black-box setting, requiring no access to the target model's parameters, and leverages the shared architectural and latent space properties of diffusion models. Our method can be seamlessly integrated into most regenerative diffusion models without retraining or optimization, offering an efficient and generalizable solution for semantic watermark forgery.



%% file: Sections/Experiments.tex
\section{Evaluation}
\label{sec:experiments}
In this section, we evaluate the watermark forgery risk for synthetic images with watermark signals. We explicitly ask if adversaries can forge the watermark on a target image and replicate to a set of cover images. The criteria for successful forgeries are threefold: (i) detectability of forged watermark and ownership of cover images, (ii) the visual perception of cover images after forgery, and (iii) the overhead of creating forged watermark on the cover images. 

\subsection{Setup}
\label{sec:setup}
\textbf{Datasets} To evaluate the performance of our watermark forgery method on various image types, we utilize three datasets: RealSR~\cite{cai2019realsr}, DRealSR~\cite{wei2020drealsr}, and CelebA~\cite{wang2021towards}. The RealSR and DRealSR datasets consist of real-world photographs, while the CelebA dataset focuses on human face images. All three datasets contain low-quality (LQ) and high-quality (HQ) image pairs, with resolutions of \(128 \times 128\) and \(512 \times 512\), respectively.

\textbf{Target and Proxy Models}
Target models are diffusion models that embed watermarks. We utilize four state of the art diffusion models: Stable Diffusion XL (SDXL)~\cite{podell2023sdxl}, 
PixArt-$\Sigma$~\cite{chen2310pixart}, FLUX.1~\cite{flux2024}, and Animagine XL~\cite{cagliostrolab2024animaginexl3} to generate watermarked images.
As our goal is to forge the watermark of the target model on any given image, we employ Stable Diffusion 2.1~\cite{rombach2022high} as a proxy model to extract the initial latent (\(\hat{z}_T^{(w)}\)) from watermarked images.

\textbf{Watermark Schemes}
We consider two representative semantic watermarking techniques: Tree Ring~\cite{treerings2023} and Gaussian Shading~\cite{yang2024gaussian} on the target models.

\textbf{Regenerative Diffusion Models} 
We consider the following six regenerative models: CtrlRegen~\cite{liu2024image},
StableSR~\cite{wang2024exploiting},
DiffBIR~\cite{lin2024diffbir},
SeeSR~\cite{wu2024seesr},
CCSR~\cite{sun2023ccsr}, and
HoliSDiP~\cite{tsao2024holisdip}. CtrlRegen is a watermark removal method through regenerating images, whereas the other models are designed to increase the image resolution. For the CelebA dataset, we additionally employ the face restoration variant of DiffBIR, fine-tuned on facial datasets and referred to as DiffBIR (Face).

\textbf{Metrics} We consider two types of metrics related to the detectability of forged watermark and the perceptual quality of forged watermark. We define three metrics for detectability: bit accuracy, the detection success rate, and the user attribution success rate for both watermark schemes. Gaussian Shading's bit accuracy is the ratio of correctly detected bits to total bits in the watermark. For TreeRing, the p-value reflects the distance between initial noise and the Tree Ring pattern. The Detection Success Rate (Dec.) is the true positive rate at a given false positive rate: 1e-3 for Gaussian Shading and 1e-2 for TreeRing. For Gaussian Shading , the User Attribution Success Rate (Attr.) is the percentage of the watermark correctly attributed to the user, considering 1000 users. 

We utilize a diverse set of image quality metrics to compare between the original image and the image with watermark forgery. PSNR and SSIM~\cite{wang2004image} are employed as full-reference fidelity measures, assessing pixel-level and structural similarity, respectively. For perceptual quality, we use reference-based metrics LPIPS~\cite{zhang2018unreasonable} and DISTS~\cite{ding2020image}. No-reference metrics CLIPIQA~\cite{wang2023exploring}, NIQE~\cite{zhang2015feature}, MUSIQ~\cite{ke2021musiq}, and MANIQA~\cite{yang2022maniqa} are adopted to provide a holistic assessment of visual fidelity.

\input{Tables/GS_Dec}
\subsection{Detecting Forged Watermarks}
The success of watermark forgery for Gaussian Shading can be evaluated from three perspectives: the bit accuracy of the forged watermark, its detectability, and the correctness of user attribution. Table~\ref{tab:detectability} summarizes the results against Gaussian Shading across four target models and three datasets. Our proposed attack seamlessly integrates with regenerative methods, and we evaluate it across seven such methods. Overall, the optimization-based Imprint baseline and our \algo with CtrlRegen consistently achieve the best or second-best performance across all three metrics. This can be attributed to the fact that both methods preserve the forged latent watermark without significant alteration. In contrast, the other six regenerative methods impose additional semantic constraints during generation, prioritizing cover image quality.
We further analyze the impact of different target models. The proposed \algo performs better on SDXL and Animagine XL than on PixArt-$\Sigma$ and FLUX.1. This performance gap can be explained by architectural differences: PixArt-$\Sigma$ and FLUX.1 adopt DiT-based designs, whereas our proxy model, SD2.1, is UNet-based. Furthermore, FLUX.1 introduces a distinct autoencoder architecture, making watermark forgery more challenging. The detectability results for forgery attacks against Tree-Ring can be found in our appendix.

\begin{figure}
    \centering
    \includegraphics[width=.95\linewidth]{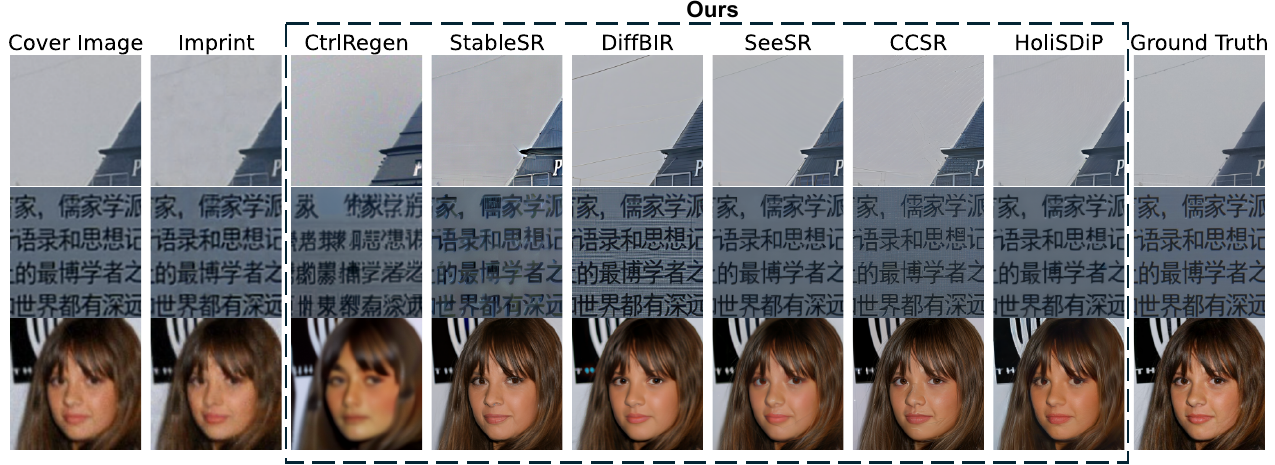}
    \caption{Example images with forged watermark using the Imprint baseline and our proposed method. The target model is SDXL with Gaussian Shading, and the proxy model used is SD 2.1.} 
    \label{fig:examples}
    \vspace{-1em}
\end{figure}

\begin{figure*}[ht]
    \centering
    \begin{minipage}{0.95\textwidth}
        \centering
        \begin{subfigure}{0.6\textwidth}
            \includegraphics[width=\textwidth]{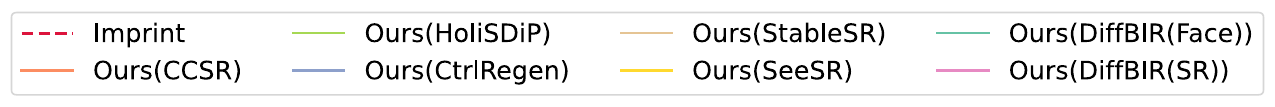}
            \label{fig:radar:legend}
        \end{subfigure}

        \begin{subfigure}{0.325\textwidth}
            \includegraphics[width=\textwidth]{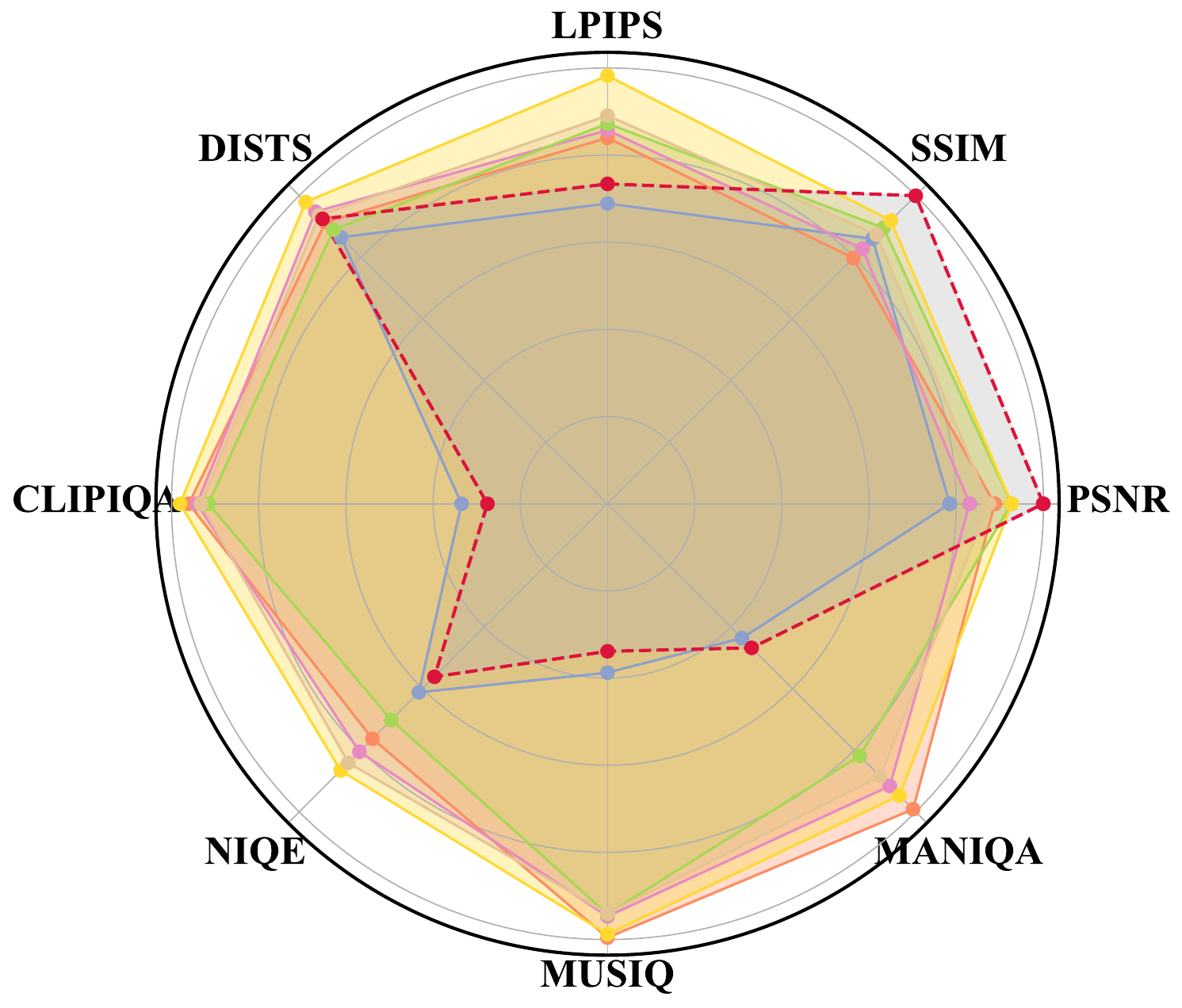}
            \caption{RealSR}
            \label{fig:radar:a}
        \end{subfigure}
        \hfill
        \begin{subfigure}{0.325\textwidth}
            \centering
            \includegraphics[width=\textwidth]{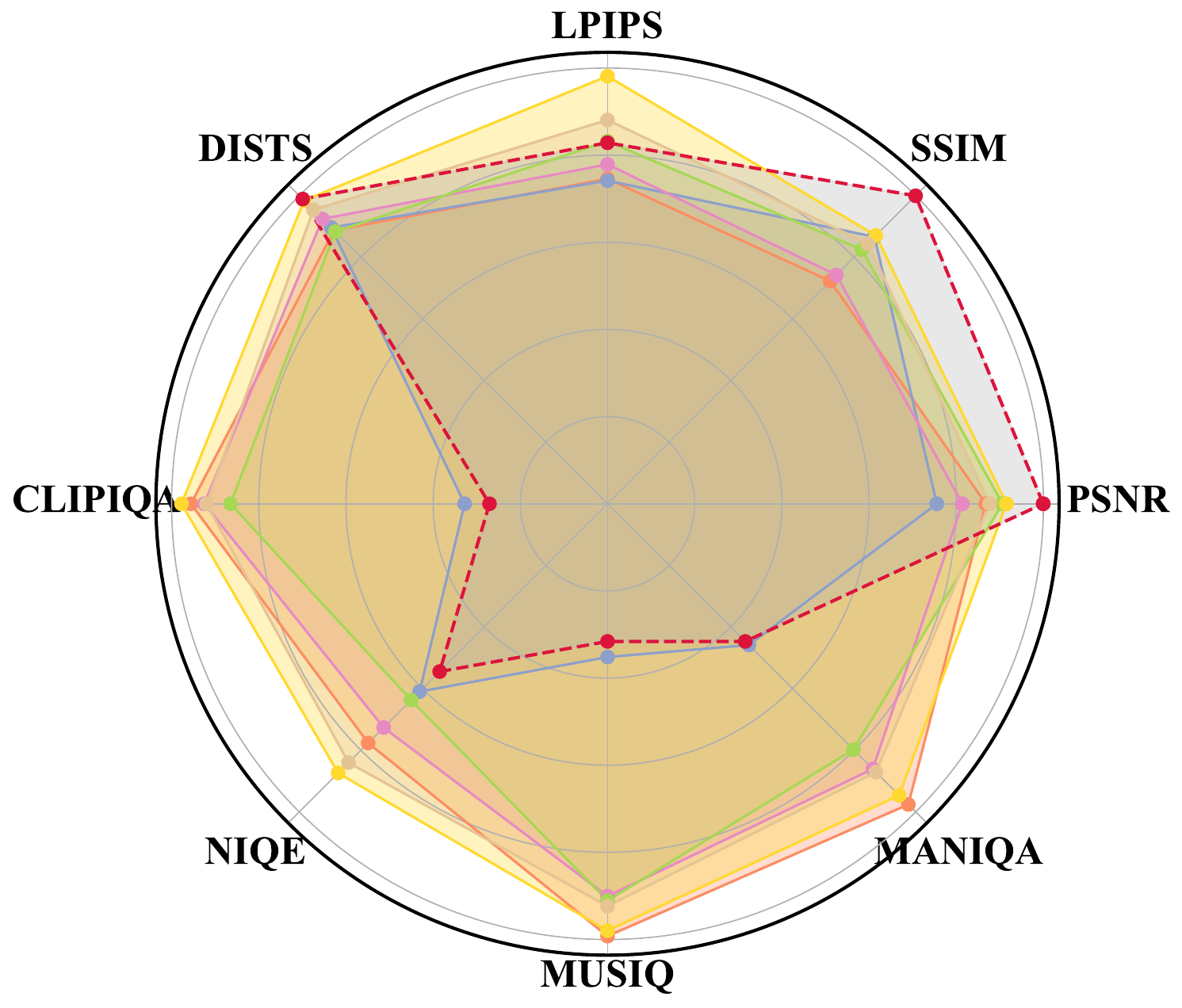}
            \caption{DRealSR}
             \label{fig:radar:b}
        \end{subfigure}
               \hfill
                \begin{subfigure}{0.325\textwidth}
            \centering
            \includegraphics[width=\textwidth]{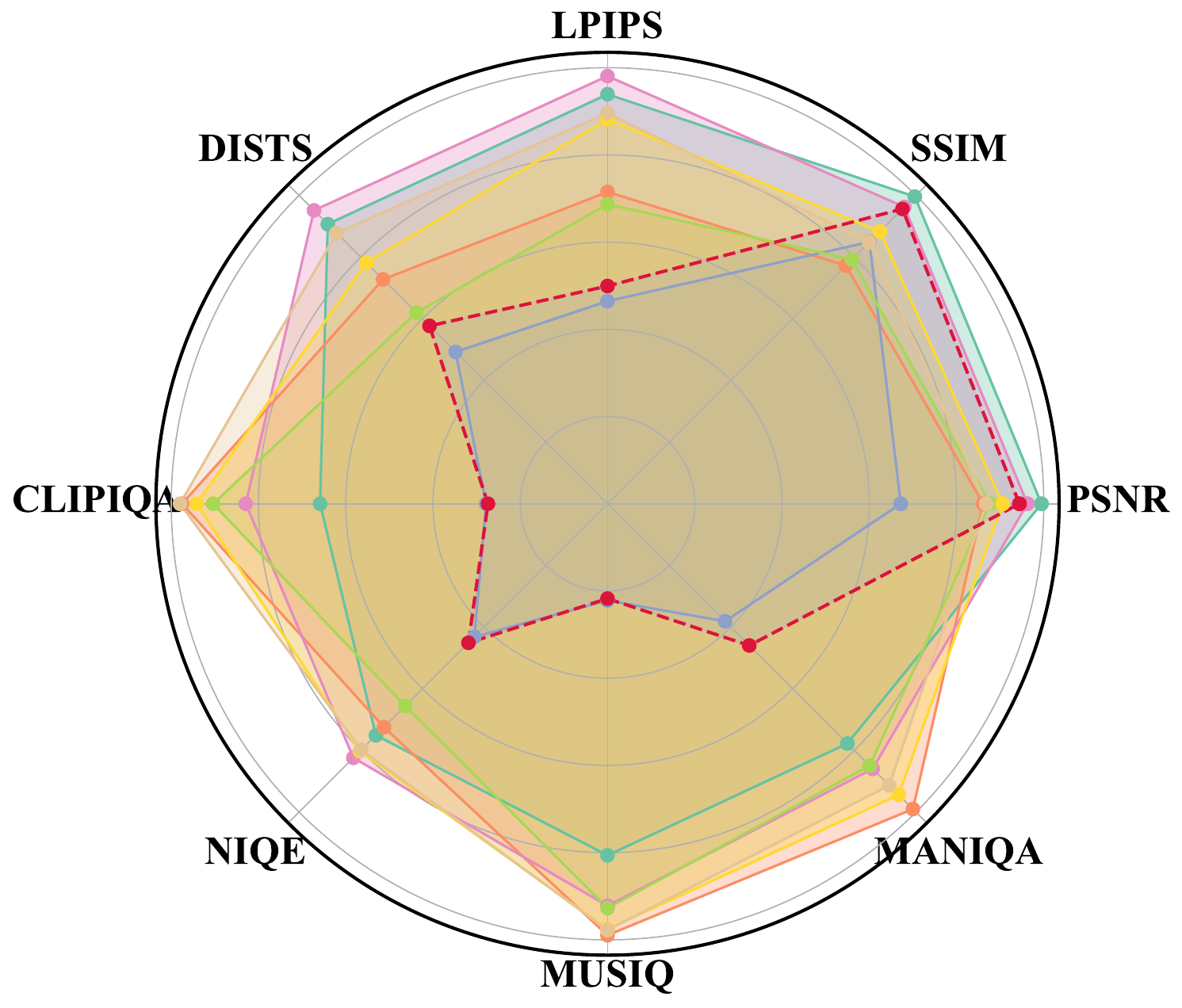}
            \caption{CelebA}
             \label{fig:radar:c}
        \end{subfigure}
        \caption{Quality of forged watermarks generated by \algo and the Imprint baseline. Metrics are averaged across different target models and watermarks. Scores like LPIPS and NIQE (where lower scores indicate better image quality) are inverted and normalized for consistency.}
        \label{fig:quality}
    \end{minipage}
\end{figure*}
\subsection{Quality of Forged Images}
One of the primary shortcomings of state-of-the-art forgery attacks is the degradation of image quality when embedding forged watermarks. In Fig.~\ref{fig:examples}, we visualize cover images with forged watermarks using Imprint and \algo, leveraging different regenerative methods. While Imprint and CtrlRegen demonstrate the best detectability, they suffer from noticeable quality degradation, in contrast to other regenerative models designed to enhance image quality. We hypothesize that the low quality of cover images with forged watermarks may fail visual inspection or bypass forgery defenses that are yet to be developed.

We summarize eight quantitative metrics for comparing the quality of cover images in Fig.~\ref{fig:quality}. A larger area covered by a forgery attack indicates better image quality and visual perception. For all three datasets, Imprint and CtrlRegen exhibit the smallest covered areas, suggesting poorer quality compared to other super-resolution-based regenerative methods. This aligns with our visual inspection results. Notably, Imprint and CtrlRegen perform worse in non-reference metrics such as CLIPIQA, MUSIQ, NIQE, and MANIQA, further confirming their inferior overall image fidelity. We argue that achieving better performance in non-reference metrics is crucial, as reference cover images may not always be available in practical scenarios.

By integrating the proposed \algo with super-resolution regenerative models, we effectively preserve the semantics of the watermarked latent, along with textual and visual guidance from the cover images. This results in a trade-off, where the detectability of the forged watermark is reduced for improved human imperceptibility, a critical factor for defending against forgery.

\subsection{Discussion on Forgery Feasibility }
\label{sec:forgery-feasibility}
\begin{wrapfigure}{R}
{.25\textwidth}
        \centering
        \vspace{-3em}
        \includegraphics[width=.25\textwidth]{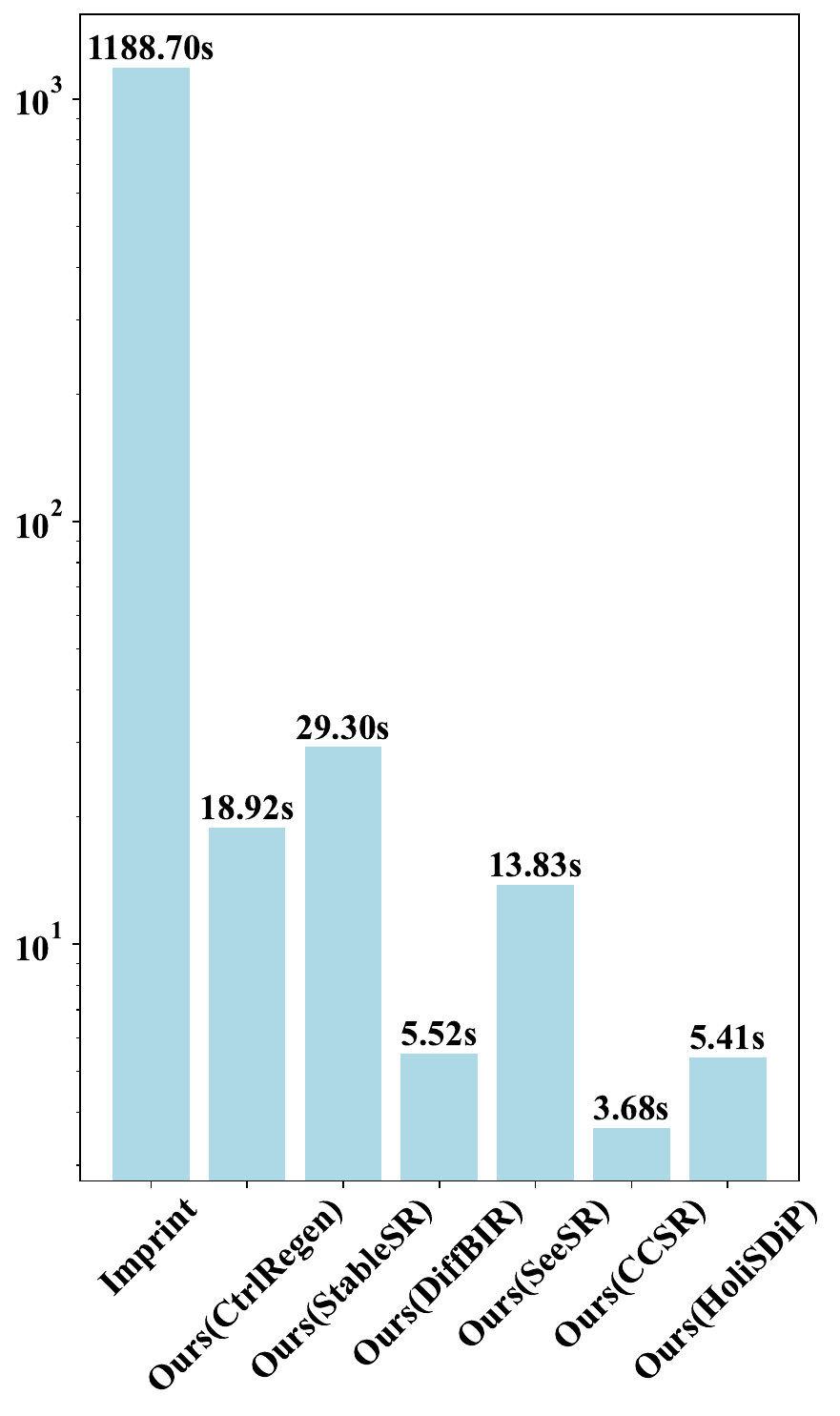}
        \caption{Average overhead required to forge watermark to a cover image.} 
                \vspace{-1em}
\label{fig:transferability}
\end{wrapfigure} 

Here, we discuss the feasibility of forgery attacks by measuring the average time required to embed a forged watermark into a single cover image, as illustrated in Fig.~\ref{fig:transferability}. Specifically, we evaluate images from the RealSR dataset with SDXL as the target model. Since Imprint relies on adversarial optimization to embed the extracted watermark into each cover image, the average generation time per image on our testbed\footnote{Hardware specifications are provided in Appendix~\ref{appx:hardware}} is approximately 1188.7 seconds. In contrast, the regenerative models used in \algo require only 3.68 to 29.3 seconds per image. This significant discrepancy highlights the prohibitive computational cost of Imprint and questions its practical applicability. Conversely, the low overhead of \algo underscores its effectiveness and practicality as a forgery attack, particularly given that the regenerative models employed are publicly available.

Our proposed watermark forgery method is both optimization-free and universal, leveraging regenerative diffusion models to perform efficient forgery. The effectiveness of the attack—measured in terms of watermark detectability, visual fidelity, and computational cost—depends heavily on the choice of the underlying generative model. For instance, while StableSR and SeeSR incur higher overhead, methods like DiffBIR, CCSR, and HoliSDiP complete the forgery in roughly 5 seconds. Notably, StableSR, DiffBIR, and SeeSR exhibit superior performance in preserving the forged watermark on the regenerated cover images (see Table~\ref{tab:detectability}). In terms of image quality, visual performance varies across datasets, as shown in Fig.~\ref{fig:quality}. Therefore, the optimal regenerative model choice is both data- and model-dependent. The low cost of executing watermark forgeries via \algo further emphasizes the urgency of developing watermarking techniques resilient to such attacks.

%% file: Tables/GS_Dec.tex
\begin{table}[]
\caption{Detectability results of our proposed \algo and baseline on various datasets and target models, evaluated against Gaussian Shading using SD 2.1 as the proxy model. The best results are highlighted in \textbf{bold}, and the second-best results are \underline{underlined}. The watermarking methods are assessed using three metrics: Bit Accuracy (Bit Acc.), Watermark Detection Success Rate (Dec.), and User Attribution Success Rate (Attr.). \label{tab:detectability}}
\centering
\resizebox{\textwidth}{!}
{\begin{tabular}{>{\centering\arraybackslash}m{2cm}>{\centering\arraybackslash}m{2cm}cccccccccc}
\toprule
\multirow{2}{*}{Target} & \multirow{2}{*}{Method} &\multirow{2}{*}{Backbone}  & \multicolumn{3}{c}{RealSR}&  \multicolumn{3}{c}{DRealSR}  & \multicolumn{3}{c}{CelebA}\\
 & & & \textbf{Bit Acc. $\uparrow$} & \textbf{Dec. $\uparrow$} & \textbf{Attr. $\uparrow$} & \textbf{Bit Acc. $\uparrow$} & \textbf{Dec. $\uparrow$} & \textbf{Attr. $\uparrow$} & \textbf{Bit Acc. $\uparrow$} & \textbf{Dec. $\uparrow$} & \textbf{Attr. $\uparrow$}   \\
\midrule
\multirow{8}{*}{SDXL} & Imprint & $-$ & \textbf{0.93} & \textbf{1.00} & \textbf{1.00} & \textbf{0.94} & \textbf{1.00} & \textbf{1.00} & \underline{0.92} & \textbf{1.00} & \textbf{1.00} \\
&\multirow{7}{*}{Ours} & CtrlRegen & \underline{0.92} & \textbf{1.00} & \textbf{1.00} & \underline{0.93} & \textbf{1.00} & \textbf{1.00} & \textbf{0.94} & \textbf{1.00} & \textbf{1.00} \\
& & StableSR & 0.85 & \textbf{1.00} & \textbf{1.00} & 0.88 & \textbf{1.00} & \textbf{1.00} & 0.90 & \textbf{1.00} & \textbf{1.00} \\
& & DiffBIR(SR) & 0.84 & \textbf{1.00} & \textbf{1.00} & 0.86 & \textbf{1.00} & \textbf{1.00} & 0.90 & \textbf{1.00} & \textbf{1.00} \\
& & DiffBIR(Face) & $-$ & $-$ &  $-$ &  $-$ &  $-$&  $-$& 0.87 & \textbf{1.00} & \textbf{1.00} \\
& & SeeSR & 0.83 & \textbf{1.00} & \textbf{1.00} & 0.86 & \textbf{1.00} & \textbf{1.00} & 0.90 & \textbf{1.00} & \textbf{1.00} \\
& & CCSR & 0.83 & \textbf{1.00} & \textbf{1.00} & 0.84 & \textbf{1.00} & \textbf{1.00} & 0.88 & \textbf{1.00} & \textbf{1.00} \\
& & HoliSDiP & 0.79 & \textbf{1.00} & \underline{0.99} & 0.81 & \underline{0.99} & \underline{0.99} & 0.88 & \textbf{1.00} & \textbf{1.00} \\
\cline{1-12}
\multirow{8}{*}{PixArt-$\Sigma$} & Imprint & $-$ & \textbf{0.77} & \textbf{1.00} & \textbf{0.97} & \textbf{0.78} & \textbf{1.00} & \textbf{0.98} & \textbf{0.77} & \textbf{1.00} & \textbf{0.97} \\
 &\multirow{7}{*}{Ours}& CtrlRegen& \underline{0.74} & \underline{0.98} & \underline{0.88} & \underline{0.75} & \underline{0.99} & \underline{0.88} & \underline{0.74} & \textbf{1.00} & \underline{0.91} \\
 & & StableSR & 0.65 & 0.74 & 0.50 & 0.69 & 0.89 & 0.58 & 0.69 & 0.94 & 0.68 \\
 & & DiffBIR(SR) & 0.67 & 0.85 & 0.62 & 0.68 & 0.91 & 0.57 & 0.71 & 0.95 & 0.83 \\
 & & DiffBIR(Face) & $-$ & $-$ &  $-$ &  $-$ &  $-$&  $-$ & 0.68 & 0.89 & 0.64 \\
& & SeeSR & 0.68 & 0.83 & 0.60 & 0.71 & 0.90 & 0.73 & 0.71 & \underline{0.98} & 0.79 \\
& & CCSR & 0.66 & 0.80 & 0.53 & 0.67 & 0.81 & 0.56 & 0.70 & 0.97 & 0.78 \\
& & HoliSDiP & 0.61 & 0.61 & 0.24 & 0.64 & 0.72 & 0.39 & 0.69 & 0.87 & 0.62 \\
\cline{1-12}
\multirow{8}{*}{FLUX.1} & Imprint & $-$& \underline{0.76} & \textbf{1.00} & \textbf{0.99} & \textbf{0.77} & \textbf{1.00} & \textbf{0.97} & \underline{0.75} & \textbf{1.00} & 0.94 \\
 &\multirow{7}{*}{Ours}& CtrlRegen& \textbf{0.76} & \textbf{1.00} & \underline{0.95} & \underline{0.76} & \underline{0.99} & \textbf{0.97} & \textbf{0.79} & \textbf{1.00} & \textbf{0.98} \\
& & StableSR & 0.66 & 0.86 & 0.62 & 0.69 & 0.94 & 0.77 & 0.71 & 0.96 & 0.89 \\
& & DiffBIR(SR) & 0.66 & 0.88 & 0.59 & 0.69 & 0.96 & 0.75 & 0.73 & \underline{0.99} & \underline{0.96} \\
& & DiffBIR(Face) & $-$ & $-$ &  $-$ &  $-$ &  $-$&  $-$ & 0.70 & 0.98 & 0.81 \\
& & SeeSR & 0.68 & \underline{0.98} & 0.62 & 0.70 & 0.96 & \underline{0.80} & 0.74 & 0.98 & 0.93 \\
& & CCSR & 0.65 & 0.81 & 0.47 & 0.66 & 0.81 & 0.54 & 0.70 & 0.97 & 0.78 \\
& & HoliSDiP & 0.62 & 0.69 & 0.31 & 0.64 & 0.72 & 0.47 & 0.70 & 0.92 & 0.81 \\
\cline{1-12}
\multirow{8}{*}{Animagine XL} & Imprint & $-$ & \textbf{0.90} & \textbf{1.00} & \textbf{1.00} & \textbf{0.92} & \textbf{1.00} & \textbf{1.00} & \underline{0.90} & \textbf{1.00} & \textbf{1.00} \\
 &\multirow{7}{*}{Ours}& CtrlRegen& \underline{0.90} & \textbf{1.00} & \textbf{1.00} & \underline{0.91} & \textbf{1.00} & \textbf{1.00} & \textbf{0.91} & \textbf{1.00} & \textbf{1.00} \\
& & StableSR & 0.83 & \textbf{1.00} & \textbf{1.00} & 0.85 & \textbf{1.00} & \textbf{1.00} & 0.87 & \textbf{1.00} & \textbf{1.00} \\
& & DiffBIR(SR) & 0.82 & \textbf{1.00} & \underline{0.99} & 0.84 & \textbf{1.00} & \underline{0.99} & 0.86 & \textbf{1.00} & \textbf{1.00} \\
& & DiffBIR(Face) & $-$ & $-$ & $-$ & $-$ & $-$ & $-$ & 0.83 & \textbf{1.00} & \textbf{1.00} \\
& & SeeSR & 0.80 & \textbf{1.00} & \textbf{1.00} & 0.82 & \textbf{1.00} & \textbf{1.00} & 0.85 & \textbf{1.00} & \textbf{1.00} \\
& & CCSR & 0.80 & \textbf{1.00} & \textbf{1.00} & 0.82 & \textbf{1.00} & \textbf{1.00} & 0.85 & \textbf{1.00} & \textbf{1.00} \\
& & HoliSDiP & 0.75 & \textbf{1.00} & \textbf{1.00} & 0.79 & \textbf{1.00} & \textbf{1.00} & 0.85 & \textbf{1.00} & \textbf{1.00} \\
\bottomrule
\end{tabular}
}

\end{table}

%% file: Sections/Conclusion.tex
\section{Conclusion}

While watermarks offer an effective mean to manage the governance of synthetic data from generative models, recent studies show that they unfortunately fall prey to forgery attacks under certain restricted circumstances. In this paper, we unveil a stronger watermark forgery attack that leverages off-the-shelf regenerative diffusion models. Our proposed \algo forgery is universally applicable to diffusion models and any type of image. We extract the watermarked latent from the target data and explore both the semantic latent as a prior and textual-visual guidance from cover images to regenerate the cover data. 
We extensively evaluate the proposed \algo on six regenerative models and 24 combinations of target models, watermarking methods and data sets, against the prior optimization-based Imprint forgery attack. Our results demonstrate that the state-of-the-art watermarks can be successfully forged from a target image to any cover image, which shows a remarkably high detection of forged watermarks and an improved image quality provided by the regenerative models. Our findings raise alarming concerns about the dependability of applying watermarks for data governance and suggest the need for designing more advanced forgery-proof watermarking methods.

\textbf{Limitations.} While \algo can be seamlessly integrated into various regenerative diffusion models, it currently requires modifying the initial latent representation to the estimated watermarked latent, and switching the sampling method to DDIM to improve forgery success rates. These adjustments, however, can degrade the overall performance of the regenerative diffusion models. Moreover, achieving different trade-offs between watermark detectability and image quality cannot be easily controlled by simply tuning the hyperparameters of a single regenerative diffusion model.

%% file: Sections/Impl_Hardware_Details.tex
\section{Implementation Details}
\label{appx:impldetail}
In this section, we provide implementation details for our target semantic watermarks, target diffusion models, baseline forgery method, regenerative diffusion models, and dataset specifications. Further details can be found in our source code.
\subsection{Watermark}
For Tree Ring~\cite{treerings2023}, we use a ring pattern with a radius of 10 and apply zero-bit watermarking. For the models used in existing work~\cite{mueller2024forgery} (SDXL, PixArt-$\Sigma$, FLUX.1), we adopt the same detection thresholds established in that study, which were derived from statistics on 5,000 watermarked images and 5,000 clean images to achieve the desired false positive rate. For Animagine XL, we follow the same procedure to compute the detection threshold by ourselves.

For Gaussian Shading~\cite{yang2024gaussian}, we follow the settings from~\cite{mueller2024forgery}. We use an encoding window of $l = 1$, with a unique random key and message generated for each image. The message length $k$ is 256, resulting in 1024 bits. The repetition factor $\rho$ is set to 64 for SD 2.1 and PixArt-$\Sigma$. For FLUX.1, the repetition factor is set to 256, since this model uses 16-channel latents, whereas the previous two models use 4-channel latents. For threshold selection to compute the true positive rate, we target a false positive rate of $10^{-3}$. In the Detection Success Rate evaluation (zero-bit scenario), we use a bit accuracy threshold of 0.5976525. For the User Attribution Success Rate evaluation (with 1, 000 users), the bit accuracy threshold is set to 0.6484375.

\subsection{Diffusion Pipeline}
All models (both \textit{Target} and \textit{Proxy}) are configured with a DDIM scheduler using 50 inference steps and a guidance scale of 7.5. For the watermark forgery experiments, we use the target models (excluding Animagine XL) to generate 100 watermarked images based on the first 100 prompts from the Stable Diffusion Prompts test set\footnote{\url{https://huggingface.co/datasets/Gustavosta/Stable-Diffusion-Prompts}}. For Animagine XL, since it performs better with structured prompts, we generated the prompts manually. The detailed prompts are available in our repository.

\begin{table}[h]
\caption{Settings of diffusion pipelines used in our experiments.}
\centering
\resizebox{\textwidth}{!}{
\begin{tabular}{l l l c c}
\toprule
\textbf{Model Name} & \textbf{Huggingface ID} & \textbf{Sampler} & \textbf{Steps} & \textbf{Guidance Scale}\\
\midrule

SD 2.1~\cite{rombach2022high} & stabilityai/stable-diffusion-2-1-base & DDIM & 50 & 7.5\\
SDXL~\cite{podell2023sdxl} & stabilityai/stable-diffusion-xl-base-1.0 & DDIM & 50 & 7.5 \\
PixArt-$\Sigma$~\cite{chen2310pixart} & PixArt-alpha/PixArt-Sigma-XL-2-512-MS & DDIM & 50 & 7.5\\
FLUX.1~\cite{flux2024} & black-forest-labs/FLUX.1-dev & DDIM & 50 & 7.5\\
Animagine XL~\cite{cagliostrolab2024animaginexl3} & cagliostrolab/animagine-xl-3.0 & DDIM & 50 & 7.5 \\
\bottomrule
\end{tabular}
}
\label{tab:diffusion_pipeline}
\end{table}
\subsection{Baseline}
We compare our method against Imprint~\cite{mueller2024forgery}.  We follow the official implementation and apply 50 steps of adversary adjustment. Parameters of the diffusion pipeline and the watermark setup are kept identical across both methods.

\subsection{Regenerative Diffusion Models}
We evaluate six regeneration models: CtrlRegen~\cite{liu2024image}, StableSR~\cite{wang2024exploiting}, DiffBIR~\cite{lin2024diffbir}, SeeSR~\cite{wu2024seesr}, CCSR~\cite{sun2023ccsr}, and HoliSDiP~\cite{tsao2024holisdip}. For each model, we replace the default sampler with DDIM and set the number of sampling steps to 50. All other hyperparameters follow the default configurations provided in their respective official repositories.
\begin{table}[h]
\caption{Settings of regenerative models used in our experiments.}
\centering
\begin{tabular}{l l l c}
\toprule
\textbf{Model Name} & \textbf{Official Repository} & \textbf{Sampler} & \textbf{Steps} \\
\midrule
CtrlRegen~\cite{liu2024image} & \url{https://github.com/yepengliu/CtrlRegen} & DDIM & 50 \\
StableSR~\cite{wang2024exploiting} & \url{https://github.com/IceClear/StableSR} & DDIM & 50 \\
DiffBIR~\cite{lin2024diffbir} & \url{https://github.com/XPixelGroup/DiffBIR} & DDIM & 50 \\
SeeSR~\cite{wu2024seesr} & \url{https://github.com/cswry/SeeSR} & DDIM & 50 \\
CCSR~\cite{sun2023ccsr} & \url{https://github.com/csslc/CCSR} & DDIM & 50 \\
HoliSDiP~\cite{tsao2024holisdip} & \url{https://github.com/liyuantsao/HoliSDiP} & DDIM & 50 \\
\bottomrule
\end{tabular}
\label{tab:regenerative_models}
\end{table}

\subsection{Datasets}
We use three datasets as cover images for watermark embedding. 

The \textbf{RealSR}~\cite{cai2019realsr} test set consists of high-resolution (HR) and low-resolution (LR) image pairs captured by two full-frame DSLR cameras (Canon 5D3 and Nikon D810). It includes 15 pairs for each camera across three scale settings, totaling 90 pairs.

The \textbf{DRealSR}~\cite{wei2020drealsr} test set comprises 93 image pairs, randomly selected at a scale of 4x from five different DSLR cameras.

The \textbf{CelebA}~\cite{wang2021towards} test set contains 3,000 CelebA-HQ images from the testing partition. The low-quality (LQ) images are generated by a degradation model using the high-quality images as input. For testing, we randomly select 100 images from this set.

\section{Hardware Details}
\label{appx:hardware}
All experiments were conducted on a remote server equipped with 8 NVIDIA A800-SXM4-80GB GPUs, 2 Intel Xeon Gold 6326 CPUs (64 cores total), and 128GB of RAM. The system runs Ubuntu 22.04 with Linux kernel 6.2 and uses CUDA 12.2 and driver version 535.86.10. Experiments described in Section~\ref{sec:forgery-feasibility} are run on a single A800 GPU. All methods were evaluated within the same batch under identical system conditions.


%% file: Sections/Quality_Table.tex
\section{Additional Experimental Results}
In this section, we present additional results for detectability against the Tree Ring watermark and detailed quality results for all watermarks and target models.

\subsection{Quantitative Results on Detectability Against Tree Ring}
Table~\ref{tab:TR_Dec} shows the detectability results for Tree Ring watermarks. It presents the watermark detection success rates for both our method and the baseline across various datasets and target models, evaluated against Tree Ring using SD 2.1 as the proxy model. The Watermark Detection Success Rate is defined as the true positive rate under a 1\% false positive rate. From the results, we observe that, for SDXL and Animagine XL, the forged watermark demonstrates better detectability. However, for PixArt-$\Sigma$ and FLUX.1, the watermarks are almost undetectable. The reason for this is that SDXL and Animagine XL show higher similarity with our target model, SD 2.1. Additionally, the lower robustness of the Tree Ring watermark also contributes to the reduced Watermark Detection Success Rate.

\subsection{Quantitative Results on Image Quality after Watermark Forgery}  
This subsection presents quantitative results for each target model (SDXL, PixArt-$\Sigma$, FLUX.1, and Animagine XL), complementing the analysis in~Fig. \ref{fig:quality}. The results are organized by the watermarking method (Tree Ring and Gaussian Shading) and the dataset (realSR, DRealSR, and CelebA), as shown in~\cref{tab:realsr_gs,tab:drealsr_gs,tab:celeba_gs,tab:drealsr_tr,tab:realsr_tr,tab:celeba_tr}
. We evaluate image quality after watermark forgery using a comprehensive set of metrics. The best results are highlighted in \textbf{bold}, and the second-best results are \underline{underlined}.

\input{Tables/TR_Dec}

\input{Sections/Addtional-quality-tables}

%% file: Tables/TR_Dec.tex
\begin{table}[H]
\centering
\caption{Detectability results of our method and baseline on various datasets and target models, evaluated against Tree Ring using SD 2.1 as the proxy model. The best results are highlighted in \textbf{bold}, and the second-best results are \underline{underlined}. The watermarking methods are assessed using two metrics: average p-value (p-value), Watermark Detection Success Rate (Dec.).}
\resizebox{0.95\textwidth}{!}{\begin{tabular}{>{\centering\arraybackslash}m{2cm}>{\centering\arraybackslash}m{2cm}ccccccc}
\toprule
\multirow{2}{*}{Target} & \multirow{2}{*}{Method} &\multirow{2}{*}{Backbone}  & \multicolumn{2}{c}{RealSR}&  \multicolumn{2}{c}{DRealSR}  & \multicolumn{2}{c}{CelebA}\\
 &  & &\textbf{p-value $\downarrow$} & \textbf{Dec. $\uparrow$} & \textbf{p-value $\downarrow$} & \textbf{Dec. $\uparrow$} & \textbf{p-value $\downarrow$} & \textbf{Dec. $\uparrow$} \\

\midrule
\multirow{8}{*}{SDXL} & Imprint & $-$&\underline{0.01} & \underline{0.97} & \underline{0.01} & \underline{0.91} & 0.03 & 0.76 \\
&\multirow{7}{*}{Ours} & CtrlRegen & \textbf{0.00} & \textbf{0.98} & \textbf{0.00} & \textbf{1.00} & \textbf{0.00} & \textbf{1.00} \\
 && StableSR & 0.01 & 0.92 & 0.01 & 0.87 & 0.02 & 0.88 \\
 && DiffBIR(SR) & 0.05 & 0.73 & 0.03 & 0.80 & 0.04 & 0.70 \\
 && DiffBIR(Face) & $-$ & $-$ & $-$& $-$ & 0.07 & 0.65 \\
 && SeeSR & 0.02 & 0.88 & 0.02 & 0.84 & 0.01 & 0.86 \\
 && CCSR & 0.02 & 0.81 & 0.01 & 0.84 & 0.01 & 0.90 \\
 && HoliSDiP & 0.04 & 0.69 & 0.02 & 0.83 & \underline{0.01} & \underline{0.94} \\
\cline{1-9}
\multirow{8}{*}{PixArt-$\Sigma$} & Imprint & $-$ & \underline{0.09} & \underline{0.51} & \underline{0.09} & \underline{0.60} & 0.13 & 0.30 \\
&\multirow{7}{*}{Ours}& CtrlRegen & \textbf{0.08} & \textbf{0.53} & \textbf{0.05} & \textbf{0.72} & \textbf{0.07} & \textbf{0.52} \\
&& StableSR & 0.21 & 0.26 & 0.16 & 0.37 & 0.18 & 0.21 \\
&& DiffBIR(SR) & 0.16 & 0.25 & 0.15 & 0.37 & 0.17 & 0.33 \\
&& DiffBIR(Face) & $-$ & $-$ & $-$& $-$  & 0.21 & 0.20 \\
&& SeeSR & 0.16 & 0.37 & 0.12 & 0.47 & \underline{0.11} & \underline{0.48} \\
&& CCSR & 0.19 & 0.32 & 0.11 & 0.41 & 0.12 & 0.37 \\
&& HoliSDiP & 0.19 & 0.23 & 0.14 & 0.32 & 0.13 & 0.45 \\
\cline{1-9}
\multirow{8}{*}{FLUX.1} & Imprint &$-$& \textbf{0.20} & \textbf{0.17} & \textbf{0.18} & \textbf{0.12} & 0.25 & 0.07 \\
&\multirow{7}{*}{Ours}& CtrlRegen & \underline{0.21} & \underline{0.13} & \underline{0.21} & \underline{0.09} & \underline{0.22} & 0.06 \\
 && StableSR & 0.27 & 0.05 & 0.22 & \underline{0.09} & 0.27 & \textbf{0.13} \\
 && DiffBIR(SR) & 0.29 & 0.04 & 0.27 & 0.06 & 0.31 & 0.02 \\
 && DiffBIR(Face) & $-$ & $-$ & $-$& $-$ & 0.34 & 0.07 \\
 && SeeSR & 0.28 & 0.09 & 0.25 & \textbf{0.12} & 0.27 & 0.04 \\
 && CCSR & 0.34 & 0.05 & 0.35 & 0.03 & 0.29 & 0.04 \\
 && HoliSDiP & 0.28 & 0.06 & 0.28 & 0.08 & \textbf{0.19} & \underline{0.12} \\
\cline{1-9}
\multirow{8}{*}{Animagine XL} & Imprint &$-$& \underline{0.03} & \underline{0.79} & \underline{0.03} & \underline{0.78} & 0.06 & 0.52 \\
&\multirow{7}{*}{Ours}& CtrlRegen & \textbf{0.01} & \textbf{0.88} & \textbf{0.01} & \textbf{0.89} & \textbf{0.02} & \textbf{0.84} \\
&& StableSR & 0.04 & 0.71 & 0.03 & 0.76 & 0.05 & 0.59 \\
&& DiffBIR(SR) & 0.09 & 0.46 & 0.09 & 0.44 & 0.09 & 0.38 \\
&& DiffBIR(Face) & $-$ & $-$ & $-$& $-$ & 0.11 & 0.30 \\
&& SeeSR & 0.06 & 0.64 & 0.05 & 0.62 & 0.05 & 0.66 \\
&& CCSR & 0.07 & 0.53 & 0.04 & 0.66 & 0.04 & 0.62 \\
&& HoliSDiP & 0.08 & 0.51 & 0.06 & 0.62 & \underline{0.04} & \underline{0.72} \\
\bottomrule
\end{tabular}
}
\label{tab:TR_Dec}
\end{table}

%% file: Sections/Addtional-quality-tables.tex
\begin{table}[h]
\centering
\caption{Image quality comparison of \algo and the Imprint baseline on RealSR images with forged Gaussian Shading watermarks.}
\resizebox{\textwidth}{!}{\begin{tabular}{>{\centering\arraybackslash}m{2cm}>{\centering\arraybackslash}m{2cm}cccccccc}
\toprule
 &  & \textbf{PSNR $\uparrow$} & \textbf{SSIM $\uparrow$} & \textbf{LPIPS $\downarrow$} & \textbf{DISTS $\downarrow$} & \textbf{CLIPIQA $\uparrow$} & \textbf{NIQE $\downarrow$} & \textbf{MUSIQ $\uparrow$} & \textbf{MANIQA $\uparrow$} \\
\textbf{Target} & \textbf{Method} &  &  &  &  &  &  &  &  \\
\midrule
\multirow{7}{*}{SDXL} & Imprint & \textbf{26.14} & \textbf{0.73} & 0.48 & 0.27 & 0.20 & 9.13 & 24.73 & 0.34 \\
 & CtrlRegen & 20.45 & 0.63 & 0.51 & 0.28 & 0.23 & 8.37 & 29.07 & 0.32 \\
 & StableSR & 22.62 & 0.64 & \underline{0.39} & \underline{0.26} & 0.67 & \textbf{5.29} & 68.99 & 0.64 \\
 & DiffBIR(SR) & 21.53 & 0.61 & 0.40 & 0.26 & 0.68 & \underline{5.78} & 69.75 & 0.66 \\
 & SeeSR & \underline{24.13} & \underline{0.67} & \textbf{0.35} & \textbf{0.25} & \underline{0.69} & 5.80 & \underline{72.36} & \underline{0.68} \\
 & CCSR & 22.99 & 0.59 & 0.41 & 0.27 & \textbf{0.70} & 6.27 & \textbf{73.00} & \textbf{0.72} \\
 & HoliSDiP & 23.62 & 0.64 & 0.42 & 0.28 & 0.69 & 6.73 & 71.25 & 0.62 \\
\cline{1-10}
\multirow{7}{*}{PixArt-$\Sigma$} & Imprint & \textbf{26.16} & \textbf{0.73} & 0.48 & \underline{0.27} & 0.20 & 9.09 & 24.68 & 0.33 \\
 & CtrlRegen & 20.68 & 0.63 & 0.52 & 0.30 & 0.24 & 8.62 & 27.57 & 0.30 \\
 & StableSR & 23.44 & 0.66 & 0.39 & 0.28 & 0.68 & 7.52 & 67.97 & 0.62 \\
 & DiffBIR(SR) & 22.16 & 0.60 & 0.42 & 0.27 & 0.67 & \underline{7.26} & 68.20 & 0.64 \\
 & SeeSR & 24.34 & 0.67 & \underline{0.37} & \textbf{0.26} & \textbf{0.71} & \textbf{6.43} & \underline{71.85} & \underline{0.68} \\
 & CCSR & 23.48 & 0.58 & 0.43 & 0.28 & \underline{0.68} & 7.32 & \textbf{72.34} & \textbf{0.70} \\
 & HoliSDiP & \underline{25.37} & \underline{0.71} & \textbf{0.37} & 0.28 & 0.61 & 8.87 & 63.93 & 0.53 \\
\cline{1-10}
\multirow{7}{*}{FLUX.1} & Imprint & \textbf{26.16} & \textbf{0.73} & 0.48 & \underline{0.27} & 0.20 & 9.14 & 24.73 & 0.33 \\
 & CtrlRegen & 20.67 & 0.63 & 0.51 & 0.30 & 0.25 & 8.19 & 27.67 & 0.31 \\
 & StableSR & 23.00 & 0.64 & \underline{0.41} & 0.28 & 0.69 & 7.19 & 68.61 & 0.64 \\
 & DiffBIR(SR) & 21.76 & 0.60 & 0.43 & 0.27 & 0.68 & 7.41 & 68.49 & 0.66 \\
 & SeeSR & \underline{24.40} & \underline{0.68} & \textbf{0.36} & \textbf{0.25} & \textbf{0.72} & \textbf{6.22} & \underline{72.10} & \underline{0.68} \\
 & CCSR & 23.58 & 0.60 & 0.41 & 0.27 & \underline{0.70} & \underline{7.10} & \textbf{72.91} & \textbf{0.72} \\
 & HoliSDiP & 23.56 & 0.59 & 0.47 & 0.30 & 0.66 & 8.45 & 69.33 & 0.58 \\
\cline{1-10}
\multirow{7}{*}{Animagine XL} & Imprint & \textbf{26.14} & \textbf{0.73} & 0.48 & 0.27 & 0.20 & 9.16 & 24.72 & 0.34 \\
 & CtrlRegen & 20.36 & 0.62 & 0.51 & 0.28 & 0.23 & 8.34 & 28.85 & 0.32 \\
 & StableSR & 22.47 & 0.62 & 0.39 & 0.26 & 0.64 & \textbf{5.13} & 68.90 & 0.64 \\
 & DiffBIR(SR) & 21.49 & 0.61 & 0.39 & \underline{0.25} & 0.68 & 5.48 & 70.05 & 0.66 \\
 & SeeSR & 24.12 & 0.67 & \textbf{0.35} & \textbf{0.25} & \textbf{0.70} & \underline{5.35} & \underline{72.12} & \underline{0.68} \\
 & CCSR & 22.89 & 0.57 & 0.42 & 0.27 & \underline{0.69} & 6.31 & \textbf{72.47} & \textbf{0.72} \\
 & HoliSDiP & \underline{24.41} & \underline{0.68} & \underline{0.38} & 0.26 & 0.68 & 5.95 & 70.49 & 0.62 \\
\bottomrule
\end{tabular}
}

\label{tab:realsr_gs}
\end{table}
\begin{table}[h]
\centering
\caption{Image quality comparison of \algo and the Imprint baseline on DRealSR images with forged Gaussian Shading watermarks.}
\resizebox{\textwidth}{!}{\begin{tabular}{>{\centering\arraybackslash}m{2cm}>{\centering\arraybackslash}m{2cm}cccccccc}
\toprule
 &  & \textbf{PSNR $\uparrow$} & \textbf{SSIM $\uparrow$} & \textbf{LPIPS $\downarrow$} & \textbf{DISTS $\downarrow$} & \textbf{CLIPIQA $\uparrow$} & \textbf{NIQE $\downarrow$} & \textbf{MUSIQ $\uparrow$} & \textbf{MANIQA $\uparrow$} \\
\textbf{Target} & \textbf{Method} &  &  &  &  &  &  &  &  \\
\midrule
\multirow{7}{*}{SDXL} & Imprint & \textbf{29.24} & \textbf{0.81} & 0.47 & \underline{0.28} & 0.20 & 10.53 & 22.04 & 0.31 \\
 & CtrlRegen & 21.91 & 0.70 & 0.51 & 0.29 & 0.22 & 9.40 & 25.13 & 0.32 \\
 & StableSR & 25.28 & 0.68 & \underline{0.44} & 0.28 & 0.68 & \textbf{6.00} & 65.80 & 0.61 \\
 & DiffBIR(SR) & 23.39 & 0.61 & 0.49 & 0.29 & 0.68 & 7.18 & 64.72 & 0.60 \\
 & SeeSR & \underline{26.65} & \underline{0.71} & \textbf{0.39} & \textbf{0.27} & \textbf{0.72} & \underline{6.25} & \underline{69.38} & \underline{0.65} \\
 & CCSR & 25.10 & 0.58 & 0.51 & 0.30 & \underline{0.71} & 6.90 & \textbf{70.14} & \textbf{0.68} \\
 & HoliSDiP & 25.71 & 0.65 & 0.49 & 0.32 & 0.67 & 7.49 & 67.74 & 0.59 \\
\cline{1-10}
\multirow{7}{*}{PixArt-$\Sigma$} & Imprint & \textbf{29.28} & \textbf{0.82} & 0.47 & \textbf{0.28} & 0.20 & 10.50 & 22.23 & 0.30 \\
 & CtrlRegen & 22.34 & 0.71 & 0.53 & 0.33 & 0.24 & 9.74 & 23.91 & 0.30 \\
 & StableSR & 26.29 & 0.71 & 0.44 & 0.29 & \underline{0.68} & \underline{8.50} & 63.59 & 0.57 \\
 & DiffBIR(SR) & 24.36 & 0.60 & 0.51 & 0.31 & 0.65 & 9.31 & 60.94 & 0.56 \\
 & SeeSR & 26.74 & 0.70 & \textbf{0.41} & \underline{0.29} & \textbf{0.70} & \textbf{7.78} & \underline{68.09} & \underline{0.63} \\
 & CCSR & 25.68 & 0.59 & 0.53 & 0.32 & 0.67 & 8.68 & \textbf{68.43} & \textbf{0.64} \\
 & HoliSDiP & \underline{28.05} & \underline{0.75} & \underline{0.41} & 0.30 & 0.55 & 16.05 & 57.22 & 0.46 \\
\cline{1-10}
\multirow{7}{*}{FLUX.1} & Imprint & \textbf{29.28} & \textbf{0.82} & 0.47 & \underline{0.28} & 0.20 & 10.45 & 22.20 & 0.30 \\
 & CtrlRegen & 22.25 & 0.71 & 0.53 & 0.32 & 0.27 & 9.09 & 24.56 & 0.31 \\
 & StableSR & 26.00 & 0.69 & \underline{0.44} & 0.29 & 0.69 & 7.75 & 64.81 & 0.59 \\
 & DiffBIR(SR) & 24.48 & 0.61 & 0.50 & 0.30 & 0.67 & 8.90 & 61.26 & 0.58 \\
 & SeeSR & \underline{27.17} & \underline{0.72} & \textbf{0.39} & \textbf{0.27} & \textbf{0.73} & \textbf{6.63} & \underline{68.46} & \underline{0.65} \\
 & CCSR & 25.79 & 0.60 & 0.52 & 0.30 & \underline{0.70} & \underline{7.69} & \textbf{69.73} & \textbf{0.66} \\
 & HoliSDiP & 25.66 & 0.59 & 0.54 & 0.33 & 0.64 & 9.83 & 64.83 & 0.55 \\
\cline{1-10}
\multirow{7}{*}{Animagine XL} & Imprint & \textbf{29.26} & \textbf{0.81} & 0.47 & \underline{0.28} & 0.20 & 10.47 & 22.17 & 0.31 \\
 & CtrlRegen & 21.96 & 0.70 & 0.51 & 0.29 & 0.22 & 9.28 & 25.09 & 0.32 \\
 & StableSR & 25.07 & 0.68 & \underline{0.44} & 0.28 & 0.64 & \textbf{5.71} & 64.85 & 0.60 \\
 & DiffBIR(SR) & 23.11 & 0.61 & 0.48 & 0.29 & 0.69 & 6.68 & 65.74 & 0.61 \\
 & SeeSR & 26.63 & 0.70 & \textbf{0.39} & \textbf{0.28} & \underline{0.70} & \underline{5.80} & \underline{68.95} & \underline{0.65} \\
 & CCSR & 25.09 & 0.58 & 0.52 & 0.31 & \textbf{0.71} & 6.51 & \textbf{70.08} & \textbf{0.68} \\
 & HoliSDiP & \underline{26.69} & \underline{0.70} & 0.44 & 0.29 & 0.66 & 6.73 & 65.77 & 0.58 \\

\bottomrule
\end{tabular}
}

\label{tab:drealsr_gs}
\end{table}

\begin{table}[h]
\centering
\caption{Image quality comparison of \algo and the Imprint baseline on CelebA images with forged Gaussian Shading watermarks.}
\resizebox{\textwidth}{!}{\begin{tabular}{>{\centering\arraybackslash}m{2cm}>{\centering\arraybackslash}m{2cm}cccccccc}
\toprule
 &  & \textbf{PSNR $\uparrow$} & \textbf{SSIM $\uparrow$} & \textbf{LPIPS $\downarrow$} & \textbf{DISTS $\downarrow$} & \textbf{CLIPIQA $\uparrow$} & \textbf{NIQE $\downarrow$} & \textbf{MUSIQ $\uparrow$} & \textbf{MANIQA $\uparrow$} \\
\textbf{Target} & \textbf{Method} &  &  &  &  &  &  &  &  \\
\midrule
\multirow{8}{*}{SDXL} & Imprint & 26.11 & 0.73 & 0.58 & 0.31 & 0.22 & 9.80 & 16.74 & 0.35 \\
 & CtrlRegen & 18.60 & 0.65 & 0.63 & 0.36 & 0.22 & 10.04 & 17.37 & 0.28 \\
 & StableSR & 23.86 & 0.63 & 0.33 & \underline{0.19} & \underline{0.80} & \textbf{4.80} & 76.47 & 0.69 \\
 & DiffBIR(SR) & \underline{26.49} & \underline{0.73} & \textbf{0.30} & \textbf{0.18} & 0.67 & 5.14 & 71.58 & 0.64 \\
 & SeeSR & 25.00 & 0.67 & 0.33 & 0.23 & 0.77 & \underline{5.04} & 76.55 & \underline{0.71} \\
 & CCSR & 23.63 & 0.58 & 0.40 & 0.24 & \textbf{0.81} & 5.60 & \textbf{77.59} & \textbf{0.74} \\
 & HoliSDiP & 23.37 & 0.57 & 0.44 & 0.30 & 0.77 & 5.95 & \underline{76.73} & 0.68 \\
 & DiffBIR(Face) & \textbf{27.36} & \textbf{0.75} & \underline{0.31} & 0.20 & 0.52 & 5.85 & 61.84 & 0.58 \\
\cline{1-10}
\multirow{8}{*}{PixArt-$\Sigma$} & Imprint & 26.14 & 0.73 & 0.58 & 0.31 & 0.22 & 9.84 & 17.01 & 0.35 \\
 & CtrlRegen & 18.66 & 0.65 & 0.62 & 0.38 & 0.23 & 10.68 & 17.06 & 0.29 \\
 & StableSR & 24.50 & 0.66 & 0.32 & 0.22 & \textbf{0.82} & 7.19 & \textbf{76.29} & 0.69 \\
 & DiffBIR(SR) & \underline{26.73} & \underline{0.73} & \textbf{0.30} & \textbf{0.20} & 0.70 & \textbf{5.81} & 72.32 & 0.65 \\
 & SeeSR & 25.35 & 0.68 & 0.34 & 0.25 & 0.77 & 6.48 & 75.01 & \underline{0.70} \\
 & CCSR & 24.03 & 0.59 & 0.42 & 0.27 & \underline{0.78} & 7.20 & \underline{76.02} & \textbf{0.73} \\
 & HoliSDiP & 26.12 & 0.68 & 0.38 & 0.29 & 0.69 & 9.39 & 64.69 & 0.56 \\
 & DiffBIR(Face) & \textbf{27.66} & \textbf{0.76} & \underline{0.31} & \underline{0.21} & 0.56 & \underline{6.02} & 64.41 & 0.59 \\
\cline{1-10}
\multirow{8}{*}{FLUX.1} & Imprint & 26.13 & 0.73 & 0.58 & 0.31 & 0.23 & 9.81 & 17.15 & 0.35 \\
 & CtrlRegen & 18.64 & 0.64 & 0.63 & 0.38 & 0.25 & 10.08 & 17.31 & 0.29 \\
 & StableSR & 24.21 & 0.65 & 0.33 & 0.22 & \textbf{0.82} & 6.76 & 75.64 & 0.69 \\
 & DiffBIR(SR) & \underline{26.76} & \underline{0.73} & \textbf{0.30} & \textbf{0.20} & 0.70 & \textbf{5.65} & 72.28 & 0.66 \\
 & SeeSR & 25.22 & 0.67 & 0.32 & 0.23 & 0.79 & \underline{5.88} & \underline{76.27} & \underline{0.71} \\
 & CCSR & 24.03 & 0.59 & 0.41 & 0.26 & \underline{0.80} & 6.83 & \textbf{77.03} & \textbf{0.75} \\
 & HoliSDiP & 23.16 & 0.52 & 0.51 & 0.33 & 0.74 & 7.51 & 72.13 & 0.65 \\
 & DiffBIR(Face) & \textbf{27.67} & \textbf{0.76} & \underline{0.31} & \underline{0.21} & 0.56 & 5.93 & 63.05 & 0.59 \\
\cline{1-10}
\multirow{8}{*}{Animagine XL} & Imprint & 26.14 & 0.73 & 0.58 & 0.31 & 0.23 & 9.79 & 16.92 & 0.35 \\
 & CtrlRegen & 18.49 & 0.65 & 0.63 & 0.35 & 0.21 & 10.21 & 17.43 & 0.29 \\
 & StableSR & 23.53 & 0.63 & 0.33 & 0.19 & 0.78 & \textbf{4.42} & \underline{76.52} & 0.68 \\
 & DiffBIR(SR) & \underline{26.54} & \underline{0.73} & \textbf{0.29} & \textbf{0.18} & 0.66 & 4.97 & 71.38 & 0.64 \\
 & SeeSR & 24.68 & 0.66 & 0.34 & 0.23 & 0.77 & \underline{4.93} & 76.44 & \underline{0.71} \\
 & CCSR & 23.63 & 0.58 & 0.39 & 0.24 & \textbf{0.82} & 5.23 & \textbf{77.97} & \textbf{0.75} \\
 & HoliSDiP & 24.38 & 0.63 & 0.39 & 0.26 & \underline{0.78} & 5.40 & 75.65 & 0.67 \\
 & DiffBIR(Face) & \textbf{27.41} & \textbf{0.75} & \underline{0.30} & \underline{0.19} & 0.53 & 5.77 & 62.16 & 0.59 \\

\bottomrule
\end{tabular}
}

\label{tab:celeba_gs}
\end{table}

\begin{table}[h]
\centering
\caption{Image quality comparison of \algo and the Imprint baseline on RealSR images with forged Tree Ring watermarks.}
\resizebox{\textwidth}{!}{\begin{tabular}{>{\centering\arraybackslash}m{2cm}>{\centering\arraybackslash}m{2cm}cccccccc}
\toprule
 &  & \textbf{PSNR $\uparrow$} & \textbf{SSIM $\uparrow$} & \textbf{LPIPS $\downarrow$} & \textbf{DISTS $\downarrow$} & \textbf{CLIPIQA $\uparrow$} & \textbf{NIQE $\downarrow$} & \textbf{MUSIQ $\uparrow$} & \textbf{MANIQA $\uparrow$} \\
\textbf{Target} & \textbf{Method} &  &  &  &  &  &  &  &  \\
\midrule
\multirow{7}{*}{SDXL} & Imprint & \textbf{26.14} & \textbf{0.73} & 0.48 & 0.27 & 0.20 & 9.15 & 24.52 & 0.34 \\
 & CtrlRegen & 20.57 & 0.63 & 0.50 & 0.28 & 0.23 & 8.24 & 28.69 & 0.31 \\
 & StableSR & 22.62 & 0.63 & \underline{0.39} & \underline{0.26} & 0.66 & \underline{5.53} & 68.54 & 0.64 \\
 & DiffBIR(SR) & 21.43 & 0.60 & 0.41 & 0.26 & 0.68 & 5.85 & 69.59 & 0.66 \\
 & SeeSR & \underline{24.01} & \underline{0.66} & \textbf{0.36} & \textbf{0.25} & \textbf{0.71} & \textbf{5.48} & \underline{72.03} & \underline{0.69} \\
 & CCSR & 22.79 & 0.57 & 0.42 & 0.27 & \underline{0.70} & 6.20 & \textbf{72.65} & \textbf{0.72} \\
 & HoliSDiP & 23.95 & 0.65 & 0.41 & 0.28 & 0.69 & 6.52 & 71.05 & 0.62 \\
\cline{1-10}
\multirow{7}{*}{PixArt-$\Sigma$} & Imprint & \textbf{26.18} & \textbf{0.73} & 0.48 & \underline{0.27} & 0.20 & 9.16 & 24.86 & 0.33 \\
 & CtrlRegen & 20.86 & 0.63 & 0.51 & 0.30 & 0.25 & 8.53 & 28.24 & 0.30 \\
 & StableSR & 23.47 & 0.65 & 0.40 & 0.27 & \underline{0.70} & 7.55 & 68.19 & 0.62 \\
 & DiffBIR(SR) & 21.91 & 0.59 & 0.44 & 0.28 & 0.68 & \underline{7.29} & 68.37 & 0.65 \\
 & SeeSR & 24.22 & 0.66 & \underline{0.38} & \textbf{0.27} & \textbf{0.71} & \textbf{6.49} & \underline{71.94} & \underline{0.67} \\
 & CCSR & 23.50 & 0.58 & 0.43 & 0.28 & 0.68 & 7.50 & \textbf{71.99} & \textbf{0.70} \\
 & HoliSDiP & \underline{25.44} & \underline{0.71} & \textbf{0.36} & 0.27 & 0.61 & 8.50 & 63.20 & 0.53 \\
\cline{1-10}
\multirow{7}{*}{FLUX.1} & Imprint & \textbf{26.15} & \textbf{0.73} & 0.48 & 0.27 & 0.20 & 9.11 & 24.76 & 0.33 \\
 & CtrlRegen & 20.75 & 0.64 & 0.50 & 0.28 & 0.26 & 8.90 & 28.20 & 0.30 \\
 & StableSR & 23.66 & 0.69 & \underline{0.35} & 0.26 & 0.68 & 7.51 & 67.58 & 0.63 \\
 & DiffBIR(SR) & 21.93 & 0.64 & 0.38 & 0.26 & 0.70 & 6.49 & 70.30 & 0.66 \\
 & SeeSR & \underline{24.74} & \underline{0.70} & \textbf{0.33} & \textbf{0.24} & \textbf{0.70} & \textbf{6.01} & \underline{71.58} & \underline{0.67} \\
 & CCSR & 24.00 & 0.64 & 0.38 & \underline{0.25} & \underline{0.70} & \underline{6.21} & \textbf{72.67} & \textbf{0.71} \\
 & HoliSDiP & 23.97 & 0.65 & 0.42 & 0.29 & 0.67 & 7.63 & 70.68 & 0.60 \\
\cline{1-10}
\multirow{7}{*}{Animagine XL} & Imprint & \textbf{26.15} & \textbf{0.73} & 0.48 & 0.27 & 0.20 & 9.17 & 24.79 & 0.34 \\
 & CtrlRegen & 20.39 & 0.63 & 0.51 & 0.28 & 0.22 & 8.45 & 28.13 & 0.32 \\
 & StableSR & 22.50 & 0.63 & 0.38 & \underline{0.25} & 0.62 & \underline{5.48} & 67.34 & 0.63 \\
 & DiffBIR(SR) & 21.26 & 0.60 & 0.40 & 0.26 & 0.68 & 5.67 & 70.15 & 0.67 \\
 & SeeSR & 24.17 & 0.67 & \textbf{0.36} & \textbf{0.25} & \underline{0.69} & \textbf{5.32} & \underline{72.22} & \underline{0.68} \\
 & CCSR & 23.01 & 0.58 & 0.40 & 0.27 & \textbf{0.69} & 6.23 & \textbf{72.43} & \textbf{0.72} \\
 & HoliSDiP & \underline{24.46} & \underline{0.69} & \underline{0.37} & 0.26 & 0.68 & 5.86 & 69.78 & 0.62 \\

\bottomrule
\end{tabular}
}

\label{tab:realsr_tr}
\end{table}

\begin{table}[h]
\centering
\caption{Image quality comparison of \algo and the Imprint baseline on DRealSR images with forged Tree Ring watermarks.}
\resizebox{\textwidth}{!}{\begin{tabular}{>{\centering\arraybackslash}m{2cm}>{\centering\arraybackslash}m{2cm}cccccccc}
\toprule
 &  & \textbf{PSNR $\uparrow$} & \textbf{SSIM $\uparrow$} & \textbf{LPIPS $\downarrow$} & \textbf{DISTS $\downarrow$} & \textbf{CLIPIQA $\uparrow$} & \textbf{NIQE $\downarrow$} & \textbf{MUSIQ $\uparrow$} & \textbf{MANIQA $\uparrow$} \\
\textbf{Target} & \textbf{Method} &  &  &  &  &  &  &  &  \\
\midrule
\multirow{7}{*}{SDXL} & Imprint & \textbf{29.25} & \textbf{0.81} & 0.47 & \textbf{0.28} & 0.20 & 10.52 & 22.05 & 0.31 \\
 & CtrlRegen & 22.15 & \underline{0.71} & 0.52 & 0.29 & 0.23 & 9.40 & 25.55 & 0.31 \\
 & StableSR & 25.38 & 0.67 & \underline{0.44} & 0.28 & 0.66 & \underline{5.93} & 65.69 & 0.61 \\
 & DiffBIR(SR) & 23.54 & 0.61 & 0.49 & 0.29 & 0.70 & 7.32 & 64.66 & 0.60 \\
 & SeeSR & \underline{26.51} & 0.70 & \textbf{0.40} & \underline{0.28} & \underline{0.71} & \textbf{5.82} & \underline{68.76} & \underline{0.65} \\
 & CCSR & 24.99 & 0.57 & 0.53 & 0.31 & \textbf{0.71} & 6.93 & \textbf{70.42} & \textbf{0.68} \\
 & HoliSDiP & 25.96 & 0.66 & 0.48 & 0.31 & 0.66 & 7.32 & 67.22 & 0.59 \\
\cline{1-10}
\multirow{7}{*}{PixArt-$\Sigma$} & Imprint & \textbf{29.33} & \textbf{0.82} & 0.47 & \textbf{0.28} & 0.19 & 10.54 & 22.09 & 0.30 \\
 & CtrlRegen & 22.40 & 0.71 & 0.54 & 0.33 & 0.24 & 9.64 & 24.14 & 0.31 \\
 & StableSR & 26.26 & 0.70 & 0.43 & 0.29 & \underline{0.68} & \underline{8.64} & 63.89 & 0.57 \\
 & DiffBIR(SR) & 24.32 & 0.59 & 0.52 & 0.32 & 0.65 & 9.63 & 60.38 & 0.57 \\
 & SeeSR & 26.94 & 0.71 & \underline{0.41} & \underline{0.29} & \textbf{0.72} & \textbf{7.44} & \underline{68.56} & \textbf{0.64} \\
 & CCSR & 25.65 & 0.60 & 0.52 & 0.32 & 0.67 & 9.00 & \textbf{69.14} & \underline{0.64} \\
 & HoliSDiP & \underline{28.19} & \underline{0.75} & \textbf{0.41} & 0.30 & 0.56 & 10.98 & 56.84 & 0.46 \\
\cline{1-10}
\multirow{7}{*}{FLUX.1} & Imprint & \textbf{29.29} & \textbf{0.81} & 0.46 & 0.28 & 0.20 & 10.47 & 22.11 & 0.31 \\
 & CtrlRegen & 22.43 & 0.71 & 0.51 & 0.30 & 0.27 & 9.81 & 24.72 & 0.31 \\
 & StableSR & 27.27 & \underline{0.76} & \underline{0.37} & \underline{0.27} & 0.68 & 8.22 & 61.25 & 0.57 \\
 & DiffBIR(SR) & 24.18 & 0.65 & 0.45 & 0.29 & 0.71 & 7.77 & 65.45 & 0.60 \\
 & SeeSR & \underline{27.56} & 0.75 & \textbf{0.36} & \textbf{0.26} & \textbf{0.71} & \textbf{6.64} & \underline{67.76} & \underline{0.64} \\
 & CCSR & 26.26 & 0.64 & 0.47 & 0.29 & \underline{0.71} & \underline{6.93} & \textbf{69.85} & \textbf{0.67} \\
 & HoliSDiP & 26.34 & 0.67 & 0.48 & 0.32 & 0.67 & 8.10 & 66.68 & 0.57 \\
\cline{1-10}
\multirow{7}{*}{Animagine XL} & Imprint & \textbf{29.28} & \textbf{0.81} & 0.47 & 0.28 & 0.20 & 10.46 & 22.07 & 0.31 \\
 & CtrlRegen & 22.04 & 0.70 & 0.52 & 0.29 & 0.22 & 9.33 & 25.43 & 0.32 \\
 & StableSR & 25.14 & 0.67 & \underline{0.43} & \textbf{0.27} & 0.62 & \textbf{5.76} & 63.53 & 0.60 \\
 & DiffBIR(SR) & 23.20 & 0.59 & 0.49 & 0.29 & 0.68 & 6.97 & 64.57 & 0.60 \\
 & SeeSR & 26.25 & 0.70 & \textbf{0.39} & \underline{0.27} & \textbf{0.70} & \underline{6.08} & \underline{68.77} & \underline{0.65} \\
 & CCSR & 25.08 & 0.57 & 0.52 & 0.31 & \underline{0.70} & 6.60 & \textbf{69.76} & \textbf{0.68} \\
 & HoliSDiP & \underline{26.82} & \underline{0.71} & 0.43 & 0.29 & 0.67 & 6.37 & 65.99 & 0.58 \\

\bottomrule
\end{tabular}
}

\label{tab:drealsr_tr}
\end{table}

\begin{table}[h]
\centering
\caption{Image quality comparison of \algo and the Imprint baseline on CelebA images with forged Tree Ring watermarks.}
\resizebox{\textwidth}{!}{\begin{tabular}{>{\centering\arraybackslash}m{2cm}>{\centering\arraybackslash}m{2cm}cccccccc}
\toprule
 &  & \textbf{PSNR $\uparrow$} & \textbf{SSIM $\uparrow$} & \textbf{LPIPS $\downarrow$} & \textbf{DISTS $\downarrow$} & \textbf{CLIPIQA $\uparrow$} & \textbf{NIQE $\downarrow$} & \textbf{MUSIQ $\uparrow$} & \textbf{MANIQA $\uparrow$} \\
\textbf{Target} & \textbf{Method} &  &  &  &  &  &  &  &  \\
\midrule
\multirow{8}{*}{SDXL} & Imprint & 26.11 & 0.73 & 0.58 & 0.31 & 0.23 & 9.78 & 16.74 & 0.35 \\
 & CtrlRegen & 18.67 & 0.65 & 0.63 & 0.36 & 0.22 & 9.87 & 16.87 & 0.28 \\
 & StableSR & 23.71 & 0.63 & 0.33 & 0.19 & \underline{0.80} & \textbf{4.78} & 76.63 & 0.68 \\
 & DiffBIR(SR) & \underline{26.46} & \underline{0.73} & \textbf{0.29} & \textbf{0.18} & 0.67 & \underline{4.97} & 71.51 & 0.64 \\
 & SeeSR & 24.65 & 0.66 & 0.34 & 0.23 & 0.78 & 5.01 & \underline{76.78} & \underline{0.72} \\
 & CCSR & 23.52 & 0.58 & 0.41 & 0.24 & \textbf{0.82} & 5.64 & \textbf{77.67} & \textbf{0.75} \\
 & HoliSDiP & 23.40 & 0.58 & 0.44 & 0.30 & 0.77 & 5.85 & 75.88 & 0.67 \\
 & DiffBIR(Face) & \textbf{27.30} & \textbf{0.75} & \underline{0.31} & \underline{0.19} & 0.53 & 5.80 & 61.68 & 0.58 \\
\cline{1-10}
\multirow{8}{*}{PixArt-$\Sigma$} & Imprint & 26.16 & 0.73 & 0.58 & 0.31 & 0.23 & 9.80 & 17.11 & 0.35 \\
 & CtrlRegen & 18.70 & 0.64 & 0.62 & 0.38 & 0.24 & 10.50 & 16.95 & 0.28 \\
 & StableSR & 24.48 & 0.66 & 0.32 & 0.22 & \textbf{0.83} & 7.13 & \underline{76.10} & 0.69 \\
 & DiffBIR(SR) & \underline{26.79} & \underline{0.73} & \textbf{0.30} & \textbf{0.20} & 0.70 & \textbf{5.68} & 72.32 & 0.65 \\
 & SeeSR & 25.19 & 0.67 & 0.34 & 0.25 & 0.79 & 6.72 & 75.41 & \underline{0.71} \\
 & CCSR & 24.01 & 0.59 & 0.43 & 0.27 & \underline{0.79} & 7.24 & \textbf{76.24} & \textbf{0.74} \\
 & HoliSDiP & 26.26 & 0.69 & 0.36 & 0.28 & 0.71 & 9.10 & 65.45 & 0.56 \\
 & DiffBIR(Face) & \textbf{27.66} & \textbf{0.76} & \underline{0.31} & \underline{0.20} & 0.57 & \underline{5.91} & 64.33 & 0.59 \\
\cline{1-10}
\multirow{8}{*}{FLUX.1} & Imprint & 26.14 & 0.73 & 0.58 & 0.31 & 0.23 & 9.78 & 16.89 & 0.35 \\
 & CtrlRegen & 18.82 & 0.65 & 0.61 & 0.36 & 0.26 & 10.68 & 17.96 & 0.28 \\
 & StableSR & 25.25 & 0.69 & \textbf{0.29} & 0.20 & \textbf{0.80} & 6.82 & 74.58 & 0.68 \\
 & DiffBIR(SR) & \underline{26.84} & \underline{0.74} & \underline{0.30} & \textbf{0.19} & 0.65 & \underline{5.73} & 70.04 & 0.64 \\
 & SeeSR & 25.49 & 0.70 & 0.32 & 0.22 & 0.75 & \textbf{5.46} & 75.21 & \underline{0.71} \\
 & CCSR & 24.56 & 0.62 & 0.37 & 0.23 & \underline{0.80} & 6.16 & \textbf{77.40} & \textbf{0.75} \\
 & HoliSDiP & 22.95 & 0.52 & 0.48 & 0.32 & 0.76 & 6.81 & \underline{75.41} & 0.70 \\
 & DiffBIR(Face) & \textbf{27.65} & \textbf{0.76} & 0.31 & \underline{0.20} & 0.52 & 6.05 & 60.65 & 0.58 \\
\cline{1-10}
\multirow{8}{*}{Animagine XL} & Imprint & 26.13 & 0.72 & 0.58 & 0.31 & 0.22 & 9.83 & 16.91 & 0.35 \\
 & CtrlRegen & 18.36 & 0.64 & 0.63 & 0.35 & 0.22 & 10.21 & 17.33 & 0.29 \\
 & StableSR & 23.23 & 0.63 & 0.32 & 0.19 & \underline{0.79} & \textbf{4.33} & 76.28 & 0.68 \\
 & DiffBIR(SR) & \underline{26.49} & \underline{0.73} & \textbf{0.29} & \textbf{0.18} & 0.66 & 4.98 & 71.39 & 0.64 \\
 & SeeSR & 24.34 & 0.65 & 0.35 & 0.23 & 0.77 & \underline{4.87} & \underline{76.74} & \underline{0.71} \\
 & CCSR & 23.56 & 0.58 & 0.39 & 0.23 & \textbf{0.82} & 5.11 & \textbf{77.96} & \textbf{0.75} \\
 & HoliSDiP & 24.60 & 0.64 & 0.38 & 0.26 & 0.77 & 5.15 & 75.65 & 0.67 \\
 & DiffBIR(Face) & \textbf{27.37} & \textbf{0.75} & \underline{0.30} & \underline{0.19} & 0.52 & 5.88 & 60.84 & 0.58 \\

\bottomrule
\end{tabular}
}

\label{tab:celeba_tr}
\end{table}

%% file: Sections/Appendix-addtional-examples.tex
\section{Additional Examples}
\begin{figure}[t]
    \centering
    \includegraphics[trim=0cm 16cm 0cm 16cm, clip, width=0.96\textwidth]{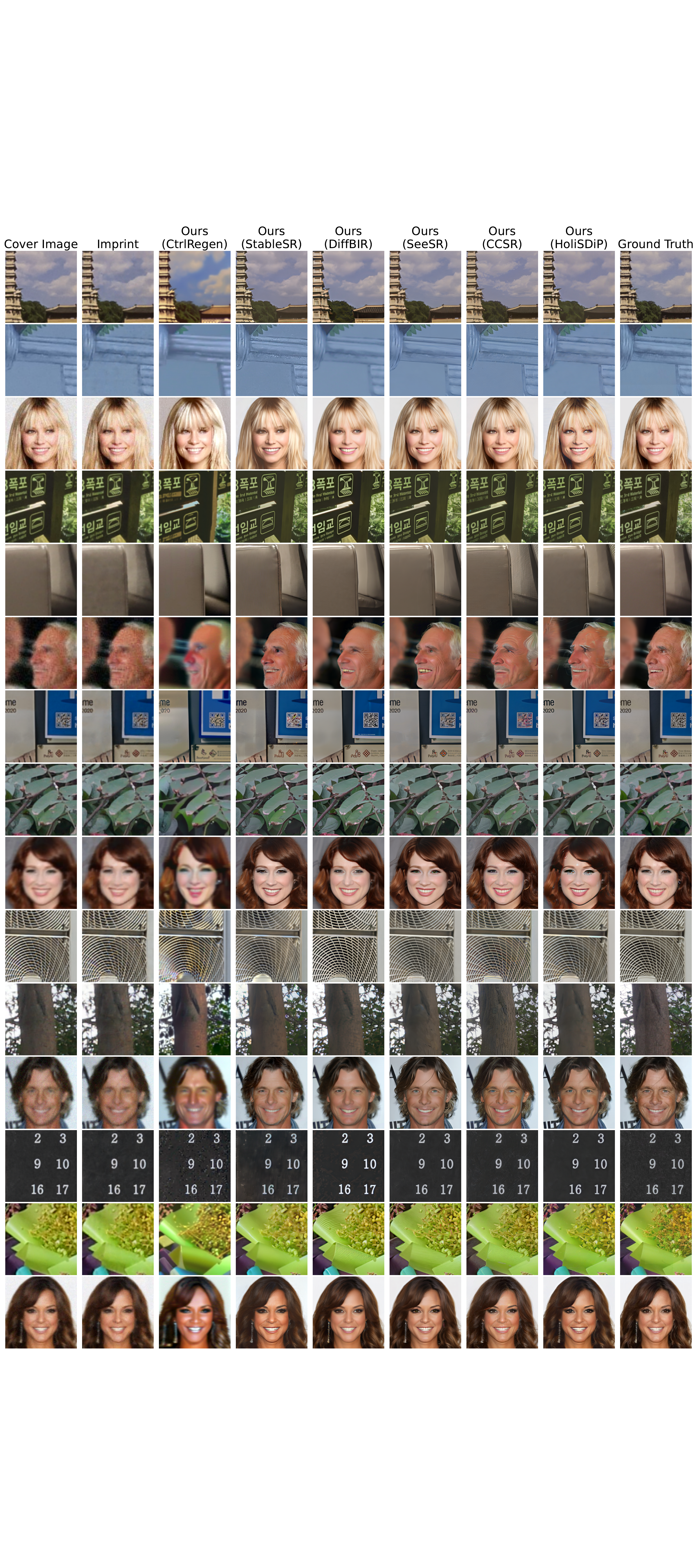}
    \caption{Additional example images with forged watermark using the Imprint baseline and our proposed method.}
    \label{fig:add-exp-1}
\end{figure}

\begin{figure}[t]
    \centering
    \includegraphics[trim=0cm 16cm 0cm 16cm, clip, width=0.96\textwidth]{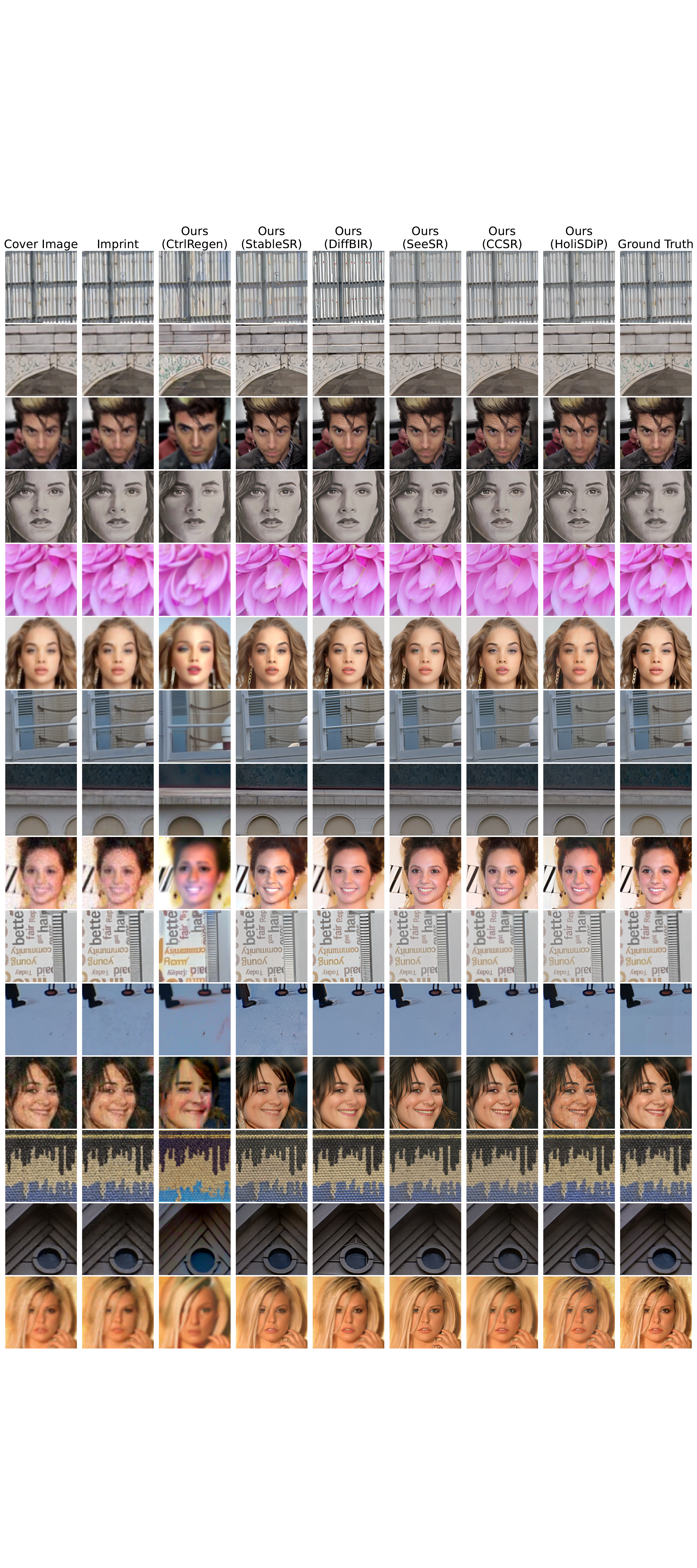}
    \caption{Additional example images with forged watermark using the Imprint baseline and our proposed method.}
    \label{fig:add-exp-2}
\end{figure}

\begin{figure}[t]
    \centering
    \includegraphics[trim=0cm 4cm 0cm 4cm, clip, width=0.95\textwidth]{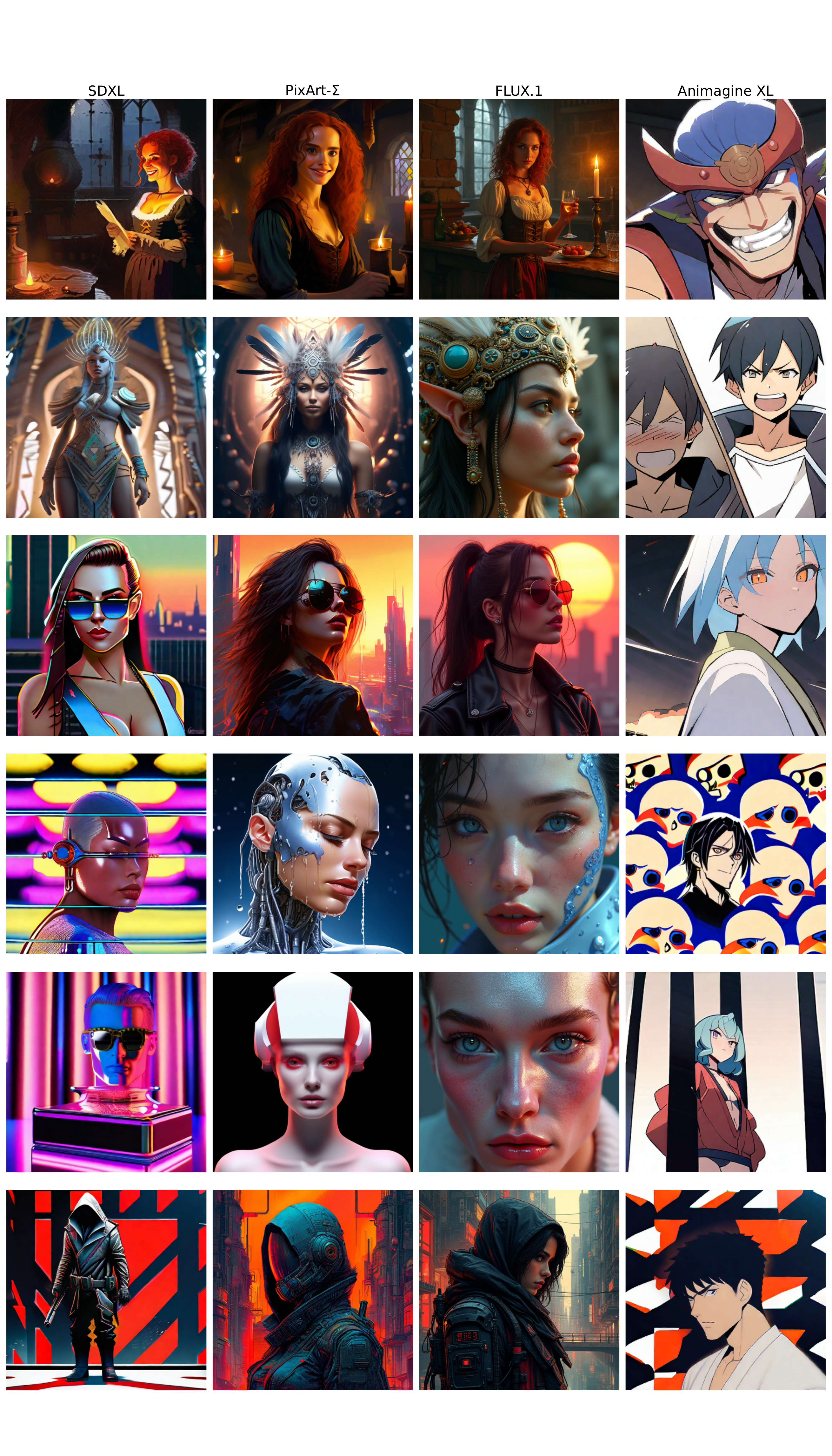}
    \caption{Images generated with the Gaussian Shading watermark using target models.}
    \label{fig:GS-exp}
\end{figure}

\begin{figure}[t]
    \centering
    \includegraphics[trim=0cm 4cm 0cm 4cm, clip, width=0.95\textwidth]{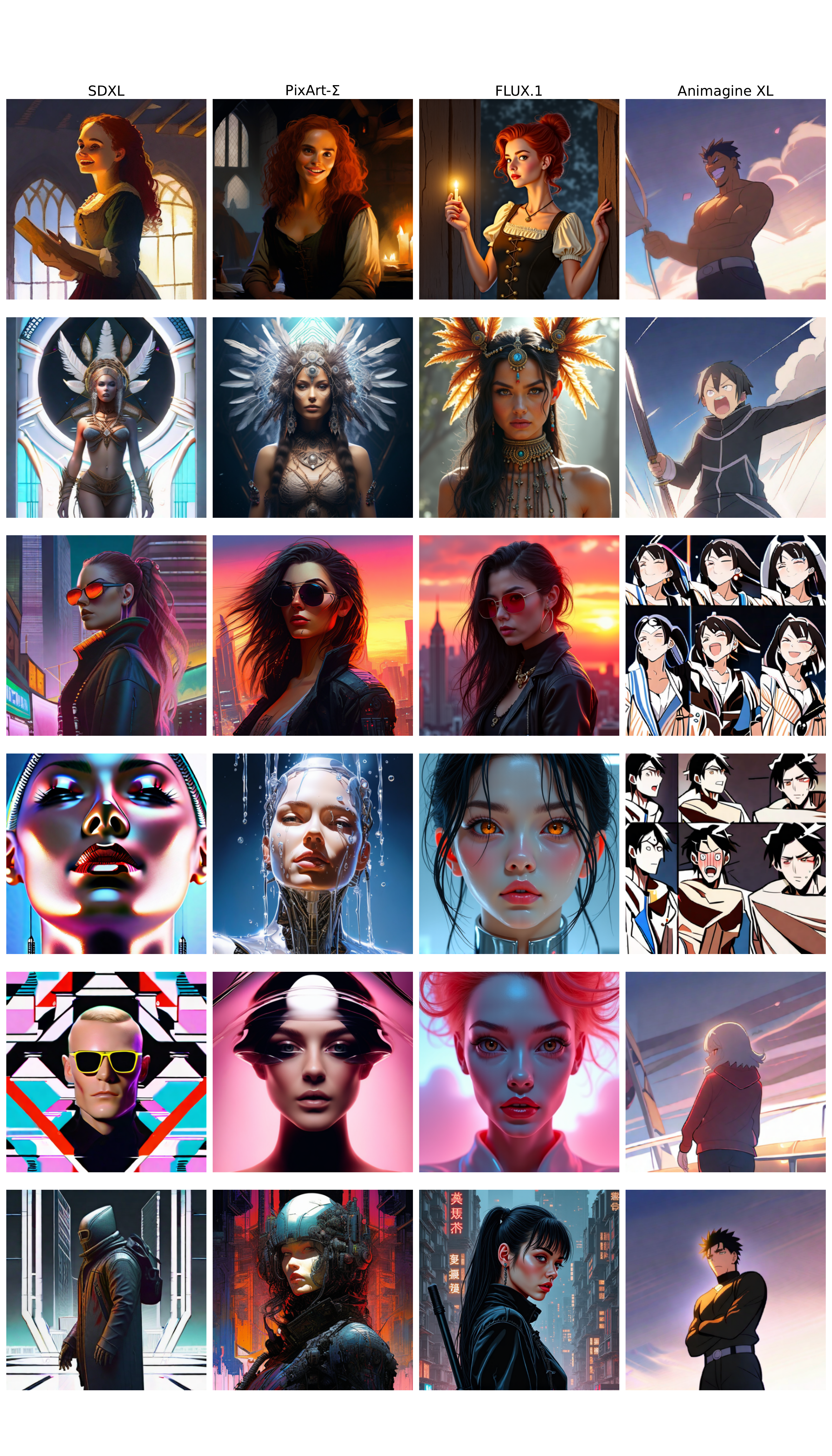}
    \caption{Images generated with the Tree Ring watermark using target models.}
    \label{fig:TR-exp}
\end{figure}

In this section, we present additional example images generated by our method. Figures~\ref{fig:add-exp-1} and~\ref {fig:add-exp-2} show more images with forged watermarks using different methods, supplementing the examples in Fig.~\ref{fig:examples}. Additionally, Figures~\ref{fig:GS-exp} and~\ref{fig:TR-exp} display watermarked images generated from different target models using Gaussian Shading and Tree Ring, respectively.

